\documentclass[onefignum,onetabnum]{siamonline220329}

\usepackage{lipsum}
\usepackage{amssymb}
\usepackage{amsfonts}
\usepackage{amsopn}
\usepackage{mathtools}
\usepackage{graphicx}
\usepackage{epstopdf}
\usepackage{algorithmic}
\usepackage{tikz}
\usepackage{pgfplots}
\usetikzlibrary{shapes.arrows, patterns, calc}
\usepackage{tikz-3dplot}
\usepackage{bbm}
\usepackage{bm}
\ifpdf
  \DeclareGraphicsExtensions{.eps,.pdf,.png,.jpg}
\else
  \DeclareGraphicsExtensions{.eps}
\fi
\usepackage{todonotes}

\definecolor{lightblue}{HTML}{a1b4c7}
\definecolor{orange}{HTML}{ea8810}
\definecolor{silver}{HTML}{b0aba8}
\definecolor{rust}{HTML}{b8420f}
\definecolor{seagreen}{HTML}{23553c}
\definecolor{joshua}{HTML}{FBDC7F}
\definecolor{darksky}{HTML}{154c79}

\colorlet{darklightblue}{lightblue!85!black}
\colorlet{darkorange}{orange!85!black}
\colorlet{lightsilver}{silver!15!white}
\colorlet{darksilver}{silver!85!black}
\colorlet{darkrust}{rust!85!black}
\colorlet{darkseagreen}{seagreen!85!black}

\colorlet{zeroborder}{darksilver}
\colorlet{zerocolor}{lightsilver}
\colorlet{nnzborder}{darksilver}
\colorlet{nnzcolor}{silver}

\colorlet{colborder}{black}
\colorlet{targetcolor}{orange}
\colorlet{selcolor}{seagreen}
\colorlet{candcolor}{lightblue}

\hypersetup{
  colorlinks=true,
  linkcolor=darkrust,
  citecolor=darkseagreen,
  urlcolor=darksilver
}

\pgfplotsset{compat=newest}
\usepgfplotslibrary{fillbetween}
\tikzset{every mark/.append style={solid}}

\newcommand*{\Reals}{\mathbb{R}}

\makeatletter
\newcommand*{\subalign}[1]{%
  \vcenter{%
    \Let@ \restore@math@cr \default@tag
    \baselineskip\fontdimen10 \scriptfont\tw@
    \advance\baselineskip\fontdimen12 \scriptfont\tw@
    \lineskip\thr@@\fontdimen8 \scriptfont\thr@@
    \lineskiplimit\lineskip
    \ialign{\hfil$\m@th\scriptstyle##$&$\m@th\scriptstyle{}##$\hfil\crcr
      #1\crcr
    }%
  }%
}
\makeatother

\newsiamremark{remark}{Remark}
\newsiamremark{hypothesis}{Hypothesis}
\newsiamremark{example}{Example}
\crefname{hypothesis}{Hypothesis}{Hypotheses}
\newsiamthm{claim}{Claim}

\headers{Cholesky factorization by conditional selection}{
S. Huan, J. Guinness, M. Katzfuss, H. Owhadi, F. Sch{\"a}fer}

\title{Sparse inverse Cholesky factorization of dense kernel matrices by greedy conditional selection}

\author{
  Stephen\;Huan\thanks{%
    Computer Science Department, Carnegie Mellon University
  } \and
  Joseph\;Guinness\thanks{%
    Department of Statistics and Data Science, Washington University in St. Louis
  } \and
  Matthias\;Katzfuss\thanks{%
    Department of Statistics, University of Wisconsin–Madison
  } \and
  Houman\;Owhadi\thanks{%
    Computing and Mathematical Sciences, Caltech, Pasadena, CA
  } \and
  Florian\;Sch{\"a}fer\thanks{%
    Georgia Institute of Technology, S1317 CODA,
    756 W Peachtree St Atlanta, GA 30332, \newline
    \email{fts@gatech.edu}, Corresponding Author
  }
}

\newcommand*{\defeq}{\coloneqq}
\newcommand*{\BigO}{\mathcal{O}}
\newcommand*{\N}{\mathcal{N}}
\newcommand*{\SpSet}{\mathcal{S}}
\newcommand*{\GP}{\mathcal{GP}}
\newcommand*{\Loss}{\mathcal{L}}
\newcommand*{\Order}{\mathcal{I}}
\newcommand*{\Reverse}{\updownarrow}
\newcommand*{\I}{I}
\newcommand*{\J}{J}
\newcommand*{\V}{V}

\renewcommand*{\vec}[1]{\bm{#1}}
\newcommand*{\Id}{\text{Id}}

\newcommand*{\CM}{\Theta}
\newcommand*{\mean}{\mu}
\newcommand*{\var}{\sigma^2}

\newcommand*{\K}{K}
\newcommand*{\Train}{\text{Tr}}
\newcommand*{\Pred}{\text{Pr}}

\DeclarePairedDelimiter{\norm}{\lVert}{\rVert}
\DeclarePairedDelimiter{\card}{\lvert}{\rvert}
\DeclareMathOperator{\diag}{diag}
\let\trace\relax
\DeclareMathOperator{\trace}{trace}
\DeclareMathOperator{\logdet}{logdet}
\DeclareMathOperator{\chol}{chol}
\DeclareMathOperator{\FRO}{FRO}

\DeclareMathOperator*{\argmin}{argmin}
\DeclareMathOperator*{\argmax}{argmax}

\DeclarePairedDelimiterX{\infdivx}[2]{(}{)}{%
  #1\;\delimsize\|\;#2%
}
\newcommand*{\KL}{\mathbb{D}_{\operatorname{KL}}\infdivx}
\DeclareMathOperator{\p}{\pi}
\DeclareMathOperator{\E}{\mathbb{E}}
\DeclareMathOperator{\Var}{\mathbb{V}ar}
\DeclareMathOperator{\Cov}{\mathbb{C}ov}
\DeclareMathOperator{\Corr}{\mathbb{C}orr}
\DeclareMathOperator{\entropy}{\mathbb{H}}
\DeclareMathOperator{\MI}{\mathbb{I}}

\ifpdf
\hypersetup{
  pdftitle={Sparse inverse Cholesky factorization of dense kernel matrices by greedy conditional selection},
  pdfauthor={S. Huan, J. Guinness, M. Katzfuss, H. Owhadi, F. Sch{\"a}fer}
}
\fi

\begin{document}

\maketitle

\begin{abstract}
  Dense kernel matrices resulting from pairwise evaluations of a
  kernel function arise naturally in machine learning and statistics.
  Previous work in constructing sparse approximate inverse Cholesky
  factors of such matrices by minimizing Kullback-Leibler divergence
  recovers the Vecchia approximation for Gaussian processes.
  These methods rely only on the geometry of the
  evaluation points to construct the sparsity pattern.
  In this work, we instead construct the sparsity pattern by leveraging
  a greedy selection algorithm that maximizes mutual information
  with target points, conditional on all points previously selected.
  For selecting \( k \) points out of \( N \), the naive time
  complexity is \( \mathcal{O}(N k^4) \), but by maintaining a
  partial Cholesky factor we reduce this to \( \mathcal{O}(N k^2) \).
  Furthermore, for multiple (\( m \)) targets we achieve a time
  complexity of \( \mathcal{O}(N k^2 + N m^2 + m^3) \), which is
  maintained in the setting of aggregated Cholesky factorization
  where a selected point need not condition every target.
  We apply the selection algorithm to image
  classification and recovery of sparse Cholesky factors.
  By minimizing Kullback-Leibler divergence, we apply the algorithm to Cholesky
  factorization, Gaussian process regression, and preconditioning the conjugate
  gradient method, improving over \( k \)-nearest neighbors selection.
\end{abstract}



\section{Introduction}

\paragraph{The problem}

Gaussian processes are widely used in spatial statistics
and geostatistics \cite{rue2005gaussian}, machine learning
through kernel methods \cite{rasmussen2006gaussian}, optimal
experimental design \cite{mutny2022experimental}, and sensor
placement \cite{krause2008nearoptimal}.
Applying Gaussian process inference to sets of \( N \) data points
requires computing with the covariance matrix \( \CM \in \Reals^{N
\times N} \) to obtain quantities such as \( \CM \vec{v} \), \(
\CM^{-1} \vec{v} \), \( \logdet(\CM) \).
For dense \( \CM \), directly computing these quantities has a
computational cost of \( \BigO(N^3) \) and a memory cost of \(
\BigO(N^2) \), which is prohibitively expensive for large \( N \).
Beyond Gaussian processes, computations with large positive-definite
matrices are required across computational mathematics,
motivating the search for faster, approximate algorithms.

\paragraph{Existing work}

Numerous methods aim to address this problem using low-rank approximations
\cite{smola2000sparse, williams2000using, fine2001efficient, bach2002kernel,
fowlkes2004spectral, chen2022randomly}, sparse approximations
\cite{furrer2006covariance,rue2005gaussian,gaedke2023parallelized},
or combinations thereof \cite{schwaighofer2002transductive,
quinonero-candela2005unifying, banerjee2008gaussian, sang2012full}.
These approximations can be viewed as imposing different
(conditional) independence structures on the Gaussian process.
Multiscale-versions of these ideas lead to wavelets \cite{beylkin1991fast,
gines1998lu}, tree codes \cite{barnes1986hierarchical,march2016askit},
panel clustering \cite{hackbusch1986complexity,hackbusch1989fast}, fast
multipole methods \cite{rokhlin1985rapid,greengard1987fast,ying2004kernel},
and various structured matrix factorizations \cite{hackbusch2000sparse,
hackbusch2002data, chandrasekaran2006fast, xia2010fast, li2012new,
ambikasaran2013mathcal, ambikasaran2015fast, ho2016hierarchical, Katzfuss2015,
Jurek2020, schafer2020compression}.
Alternatives are based on fast Fourier transforms \cite{graham2018analysis,
stein2002fast} or random feature maps \cite{rahimi2007random}.
Preconditioned iterative methods based on randomized approximations
using exact \cite{wang2019exact} or approximate matrix-vector products
\cite{meanti2020kernel,abedsoltan2023large} are popular in machine learning.

\paragraph{Vecchia approximation, Kaporin's factorization, and KL minimization}

Vecchia approximates the likelihood function of a Gaussian distribution
by decomposing it into a product of univariate conditional densities,
each of which depends on a subset of the previously ordered variables
\cite{vecchia1988estimation}.
Independently from Vecchia, Kaporin derived a closed-form expression of
the approximate inverse Cholesky preconditioner of a p.s.d. matrix that
minimizes, subject to a sparsity constraint, the \( K \)-condition number
of the preconditioned system \cite{kaporin1990alternative}.
Referring to this approximation as ``factorized sparse approximate inverse
(FSAI),'' Yeremin et al. show that Kaporin's inverse Cholesky factor also
minimizes a Frobenius-norm error metric, subject to a diagonal scaling
constraint \cite{yeremin2000factorized}.
Vecchia's likelihood approximation was later observed to
implicitly compute a sparse approximate inverse Cholesky factor
of the covariance matrix \cite{Datta2016, katzfuss2021general}.
The closed-form expression of this inverse Cholesky
factor coincides with that derived by Kaporin.
Independently from the above works, Sch{\"a}fer et al. compute sparse inverse
Cholesky factors that are optimal in Kullback-Leibler (KL) divergence
\cite{schafer2021sparse}, again recovering the formula derived by Kaporin.
The KL divergence is also used to compute Knothe-Rosenblatt transport
maps \cite{marzouk2016introduction}, generalizing Cholesky factors to
non-Gaussian distributions while preserving triangularity and sparsity
\cite{spantini2018inference}.
A method for the Bayesian estimation of such transport
maps is proposed by \cite{katzfuss2022scalable}.
A common feature of the above methods is that the columns of the
Cholesky factors can be computed independently and in parallel.

\paragraph{Ordering and sparsity selection by geometry}

\begin{figure}[t]
  \centering
  \begin{tikzpicture}[baseline]
  \begin{axis}[
    width={0.49\linewidth},
    grid={major},
    xmin=-1.1, xmax=1.1, ymin=-1.1, ymax=1.1, zmin=-0.1, zmax=1,
  ]
  \addplot3 [only marks, orange]
    table {figures/screening/data/matern_uncond_points.csv};
  \addplot3 [mesh, lightblue]
    table {figures/screening/data/matern_uncond.csv};
  \end{axis}
\end{tikzpicture}%
  \begin{tikzpicture}[baseline]
  \begin{axis}[
    width={0.49\linewidth},
    grid={major},
    xmin=-1.1, xmax=1.1, ymin=-1.1, ymax=1.1, zmin=-0.1, zmax=1,
  ]
  \addplot3 [only marks, orange]
    table {figures/screening/data/matern_cond_points.csv};
  \addplot3 [mesh, seagreen]
    table {figures/screening/data/matern_cond.csv};
  \end{axis}
\end{tikzpicture}
  \caption{%
    An illustration of the screening effect with the Mat{\'e}rn kernel with
    length scale \( \ell = 1 \) and smoothness \( \nu = 1/2 \).
    The first panel shows the \textcolor{lightblue}{unconditional
    correlation} with the point at (0, 0).
    The second panel shows the \textcolor{seagreen}{conditional correlation}
    after conditioning on the four points in \textcolor{orange}{orange}.
  }
  \label{fig:screening}
\end{figure}
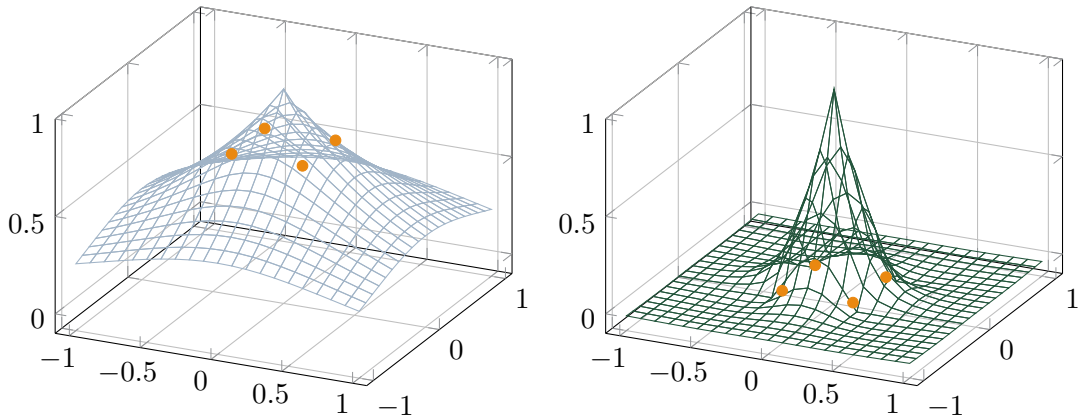

The approximation quality of the above methods depends on the ordering of the
rows and columns of the input matrix and the sparsity pattern of the factor.
Vecchia originally proposed ordering points
lexicographically \cite{vecchia1988estimation}.
The work on FSAI emphasized classical orderings from sparse linear
algebra such as (reverse) Cuthill-McKee, minimum degree, nested
dissection, and red-black ordering \cite{benzi2000orderings}.
Guinness empirically studied the effect of various orderings on
Vecchia's likelihood approximation \cite{guinness2018permutation}.
He found a random ordering and a space-filling
maximum-minimum distance (maximin) ordering to perform well.
This observation can be explained by the \emph{screening effect}, by
which conditional on points near the point of interest, far away points
are almost conditionally independent for many popular kernel functions
\cite{stein2002screening, stein20112010} (see \cref{fig:screening}).
Cholesky factorization is closely related to numerical homogenization
and operator-adapted wavelets \cite{owhadi2019operatoradapted}.
Sch{\"a}fer et al.\ exploit this connection to prove exponential decay of
Cholesky factors of covariance and precision matrices (the latter being
inverses of the former) arising from elliptic PDEs, when developed in a
multiresolution basis \cite{schafer2020compression}. This setting includes
the maximin ordering as a simplistic multiresolution basis akin to the
``lazy wavelet'' \cite{sweldens1996lifting} and thus provides a rigorous
proof of the screening effect.
The maximin ordering has since been used by \cite{schafer2021sparse,
katzfuss2021general, kang2021correlationbased, katzfuss2022scalable}.
Motivated by the screening effect, the sparsity set is often formed by
selecting the closest points by Euclidean distance \cite{vecchia1988estimation,
schafer2020compression, schafer2021sparse, katzfuss2022scalable}.
To lower the computational cost, different authors have proposed
heuristics to analyze the sparsity set and identify opportunities to
group similar points into supernodes \cite{stein2004approximating,
ferronato2015novel, guinness2018permutation}.
Sch{\"a}fer et al.\ devise a geometric grouping algorithm that allows to
provably compute an \( \epsilon \)-accurate inverse-Cholesky factor in
time complexity \( \BigO(N \log^{2d} (N/\epsilon)) \) using \( \BigO(N
\log^{ d} (N/\epsilon)) \) nonzero entries of the covariance matrix
\cite{schafer2021sparse}.
Methods based on the screening effect have improved the state-of-the-art
for solving elliptic differential and integral equations
\cite{schafer2020compression, schafer2021sparse, chen2021multiscale}.
However, selecting the sparsity set only based on distance
ignores the possible redundancy of the selected points.
The proposed work addresses this limitation.

\paragraph{Conditional selection}

Instead of adding points to the sparsity pattern by distance, we propose
greedily selecting points that maximize mutual information with the
point of interest, conditional on all points previously selected.
The machine learning community has long developed similar algorithms that
greedily optimize information-theoretic objectives in the context of
sparse Gaussian process inference \cite{smola2000sparse, herbrich2002fast,
seeger2003fast}.
Similar algorithms have also been developed in the context of
sensor placement \cite{krause2008nearoptimal, clark2018greedy}
and experimental design \cite{mutny2022experimental}, where it is
assumed the target phenomenon is modeled by a Gaussian process or
is otherwise linearly dependent on the selected measurements.
However, these works often focus on global approximation of the entire process,
e.g., through sparse approximation of the likelihood or covariance matrix
\cite{liu2020when, chalupka2012framework, quinonero-candela2005unifying}.
In contrast, \cite{wada2013gaussian} uses inference \emph{directed} towards
a point of interest, selecting the active (sparsity) set by the kernel
function itself like the later work \cite{kang2021correlationbased};
\cite{gramacy2014local} and the follow-up work \cite{gramacy2015speeding}
use the more sophisticated active learning Cohn (ALC) objective, yielding
an algorithm equivalent to ours for a single point of interest.
Our proposed algorithm can also be viewed as a variant of orthogonal matching
pursuit (OMP) \cite{tropp2007signal, tropp2006algorithms}, a workhorse
algorithm in compressive sensing, which seeks to approximate a target signal
as the sparse linear combination from a given collection of signals.

\paragraph{Main results}

Our main contribution is a selection algorithm that
greedily maximizes mutual information with point(s) of
interest, conditional on all points previously selected.
We use this algorithm to select the sparsity pattern of sparse approximate
Cholesky factors of precision matrices in the KL-minimization framework of
\cite{schafer2021sparse}, improving the approximation accuracy attainable
with a given number of nonzero entries.
Our method extends kernel-based selection \cite{wada2013gaussian,
Katzfuss2020, kang2021correlationbased} to account for conditioning.
It also extends directed Gaussian process regression \cite{gramacy2014local,
gramacy2015speeding} by simultaneously targeting multiple prediction
points and providing a global approximation of the Gaussian process.
For a single target point, naive computation of the
mutual information criterion has time complexity \(
\BigO(N k^4) \) to select \( k \) points out of \( N \).
By maintaining a partial Cholesky factor, we
reduce the complexity to \( \BigO(N k^2) \).
We extend the algorithm to maximize mutual information with
\emph{multiple} targets, re-using the same selections across
multiple targets for efficiency while maintaining accuracy.
For \( m \) target points, we achieve a time
complexity of \( \BigO(N k^2 + N m^2 + m^3) \).
If \( m \approx k \), this is \( m \) times
faster than the single-target algorithm.
In the setting of aggregated (or supernodal) Cholesky factorization, where
the sparsity patterns of multiple columns are determined simultaneously,
a candidate entry may only condition a \emph{subset} of the targets.
By rank-one downdating\footnote{%
  Given \( L = \chol(\CM) \), a \emph{rank-one downdate} of
  \( \CM \) by a vector \( \vec{u} \) efficiently computes
  \( \chol(\CM - \vec{u} \vec{u}^{\top}) \) from \( L \).
} of Cholesky factors, we capture this structure
at the same time complexity for multiple targets.
Finally, we show how to adaptively determine the number of nonzeros
per column in order to minimize the overall KL divergence by
maintaining a global priority queue shared between all columns.

\paragraph{Outline}

This paper is organized as follows.
In \cref{sec:chol}, we show how minimizing KL divergence to compute sparse
Cholesky factors reduces to solving independent regression problems.
In \cref{sec:select}, we develop greedy algorithms to select
the sparsity pattern independently for each regression problem.
In \cref{sec:chol_select}, we combine the greedy selection algorithm with
KL minimization to yield algorithms for sparse Cholesky factorization.
In \cref{sec:extensions}, we extend these results to
adjacent and nonadjacent aggregated factorization.
In \cref{sec:experiments}, we present numerical experiments applying
our method to Cholesky factorization, Gaussian process regression,
preconditioning with the conjugate gradient, image classification,
and recovery of \textit{a priori} sparse Cholesky factors.
In \cref{sec:conclusion}, we summarize our results.
Proofs and algorithmic details are provided
in the appendix and supplementary material.

\begin{figure}[t]
  \centering
  \includegraphics{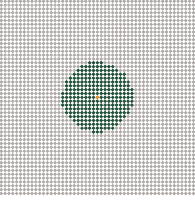}%
  \quad
  \includegraphics{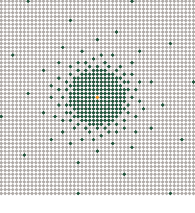}%
  \quad
  \includegraphics{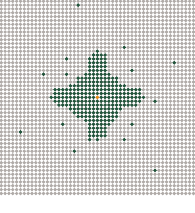}%
  \quad
  \includegraphics{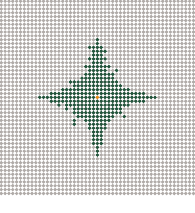}%
  \caption{%
    The first panel shows selecting the \( k \)-\textcolor{seagreen}{nearest
    neighbors} to the \textcolor{orange}{center point} out of a dense grid
    of \( 51 \times 51 \) \textcolor{silver}{candidates} for \(k = 265 \).
    The next panels show selection by greedily maximizing
    conditional mutual information with the center point for a Mat{\'e}rn
    kernel with length scale \( \ell = 1 \) and increasing smoothness \(
    \nu \), from right to left: \( \nu = 1/2, 3/2, 5/2 \).
    The full selections are found at \url{https://youtu.be/lyJf3S5ThjQ}.
  }
  \label{fig:cond_select}
\end{figure}

\section{Sparse Cholesky factorization by KL-minimization}
\label{sec:chol}

Let \( \Theta \in \Reals^{N \times N} \) be a symmetric
positive-definite matrix; we view \( \Theta \) as the
covariance matrix of a multivariate Gaussian random variable.
We say that a function \( f(\vec{x}) \) is distributed according to
a Gaussian process prior with mean function \( \mean(\vec{x}) \) and
covariance function or kernel function \( \K(\vec{x}, \vec{x}') \), which
we will denote as \( f(\vec{x}) \sim \GP(\mean(\vec{x}), \K(\vec{x},
\vec{x}')) \), if for any finite set of points \( X = \{ \vec{x}_i \}_{i
= 1}^N \), \( f(X) \sim \N(\vec{\mean}, \CM) \), where \( \mean_i =
\mu(\vec{x}_i) \) and \( \CM_{i, j} = \K(\vec{x}_i, \vec{x}_j) \).

In many applications of Gaussian processes, we
wish to infer unknown data given known data.
Given the training dataset \( \mathcal{D} = \{ (\vec{x}_i, y_i) \}_{i = 1}^N
\), where the inputs \( \vec{x}_i \in \Reals^D \) are collected in the matrix
\( X_\Train = [\vec{x}_1, \dotsc, \vec{x}_N]^{\top} \in \Reals^{N \times D} \)
and the measurements \( y_i = f(\vec{x}_i) \) at those points are collected
in the vector \( \vec{y}_\Train = [y_1, \dotsc, y_N]^{\top} \in \Reals^N \),
we wish to predict the values at \( m \) new points \( X_\Pred \in \Reals^{m
\times D} \) for which \( \vec{y}_\Pred \in \Reals^m \) is unknown.
We assume the function \( f(\vec{x}) \) that maps input points to their
outputs is distributed according to a Gaussian process with zero mean
function, \( f(\vec{x}) \sim \GP(\vec{0}, \K(\vec{x}, \vec{x}')) \).
From the distribution of \( f(\vec{x}) \), the joint distribution
of training and testing data \( \vec{y} \) has covariance
\(
  \CM =
  \begin{pmatrix}
    \CM_{\Train, \Train} & \CM_{\Train, \Pred} \\
    \CM_{\Pred, \Train} & \CM_{\Pred, \Pred}
  \end{pmatrix}
\)
where \( \CM_{\I, \J} \defeq \K(X_\I, X_\J) \) for index sets \( \I, \J \).
In order to make predictions at \( X_\Pred \), we condition the desired
prediction \( \vec{y}_\Pred \) on the known data \( \vec{y}_\Train \).
For Gaussian processes, the closed-form posterior distribution is
\begin{align}
  \label{eq:cond_mean}
  \E[\vec{y}_\Pred \mid \vec{y}_\Train] &=
    \vec{\mean}_\Pred +
    \CM_{\Pred, \Train} \CM_{\Train, \Train}^{-1}
    (\vec{y}_\Train - \vec{\mean}_\Train), \\
  \label{eq:cond_cov}
  \Cov[\vec{y}_\Pred \mid \vec{y}_\Train] &=
    \CM_{\Pred, \Pred} -
    \CM_{\Pred, \Train} \CM_{\Train, \Train}^{-1}
    \CM_{\Train, \Pred},
 \shortintertext{
    where \( \CM_{\Train, \Train}^{-1} \defeq (\CM_{\Train, \Train})^{-1} \).
    To denote the covariance between the variables in index
    sets \( \I \) and \( \J \), conditional on the variables
    in the index sets \( \V_1, \V_2, \dotsc, \V_n \) we write
  }
  \label{eq:cond_cov_notation}
  \CM_{I, J \mid \V_1, \V_2, \dotsc, \V_n} &\defeq
    \Cov[\vec{y}_\I, \vec{y}_\J \mid
         \vec{y}_{\V_1 \cup \V_2 \cup \dotsb \cup \V_n}].
  \shortintertext{
    We recursively compute \cref{eq:cond_cov_notation}; let \( W = \bigcup_{i
    = 1}^{n - 1} V_i \) and by the quotient rule of Schur complements,
  }
  \label{eq:quotient_rule}
  \CM_{\I, \J \mid \V_1, \dotsc, \V_n} &=
    \CM_{\I,   \J   \mid W} -
    \CM_{\I,   \V_n \mid W}
    \CM_{\V_n, \V_n \mid W}^{-1}
    \CM_{\V_n, \J   \mid W}.
\end{align}

Calculating the posterior mean \cref{eq:cond_mean} and
covariance \cref{eq:cond_cov} requires inverting the training
covariance matrix, usually by means of Cholesky factorization.
The time complexity of computing the Cholesky factorization
is \( \BigO(N^3) \), which is prohibitive for large \( N \).
Thus, we aim to compute \emph{sparse} approximate Cholesky factors.

\subsection{Vecchia approximation}
\label{subsec:vecchia}

The Vecchia approximation for Gaussian processes
\cite{vecchia1988estimation} can be viewed as computing
sparse approximate inverse Cholesky factors of \( \CM \).
The joint likelihood \( \p \) can be decomposed as
\begin{align}
  \label{eq:joint}
  \p(\vec{y}) &= \p(y_1) \p(y_2 \mid y_1) \p(y_3 \mid y_1, y_2) \dotsm
    \p(y_N \mid y_1, y_2, \dotsc, y_{N - 1}).
  \shortintertext{
    The key assumption is that many of the conditioning points are redundant.
    Letting \( i_1, \dotsc, i_N \) denote an ordering of the points and
    \( s_k \) the indices of points that condition the \( k \)-th point
    in the ordering, the Vecchia approximation replaces \cref{eq:joint}
    by the sparse approximation
  }
  \label{eq:vecchia}
  \p(\vec{y}) &\approx \p(y_{i_1}) \p(y_{i_2} \mid y_{s_2})
    \p(y_{i_3} \mid y_{s_3}) \dotsm \p(y_{i_N} \mid y_{s_N}).
\end{align}
The precision matrix of the resulting approximate density \cref{eq:vecchia}
has a sparse Cholesky factor in the sense that the \( i \)-th column of the
factor has the sparsity pattern \( s_i \) when written in the given elimination
ordering \cite{katzfuss2021general}.
Another way to recover this Cholesky factor for a fixed elimination ordering
\( \prec \) and lower triangular sparsity pattern \( S \defeq \{ (i, j) : i
\in s_j, i \succeq j \} \) is to specify a functional criterion \( \Loss:
\Reals^{N \times N} \to \Reals \) defining an optimization problem over
candidate matrices satisfying the sparsity pattern, \( \SpSet \defeq \{ M
\in \Reals^{N \times N} : M_{i, j} \neq 0 \Rightarrow (i, j) \in S \} \).
\begin{align}
  \label{eq:generic_obj}
  L &\defeq \argmin_{\hat{L} \in \SpSet} \Loss(\hat{L})
\end{align}
Functionals include the Kaporin condition number \( (\trace(L \CM
L^{\top})/N)^N/\det(L \CM L^{\top}) \) \cite{kaporin1990alternative}, the
Frobenius norm \( \norm{\Id - L \chol(\CM)}_{\FRO} \) additionally subject to
the constraint \( \diag(L \CM L^{\top}) = 1 \) \cite{yeremin2000factorized},
and the Kullback-Leiber (KL) divergence \( \KL{\N(\vec{0}, \CM)} {\N(\vec{0},
(L L^{\top})^{-1})} \) \cite{schafer2021sparse}.
These three choices all recover the Vecchia approximation.

As observed in \cite{rue2005gaussian, schafer2021sparse}, factors of the
precision matrix are often much sparser than factors of the covariance
matrix, because the precision encodes conditional independence while the
covariance encodes marginal independence.
The same phenomenon is observed by \cite{spantini2018inference}
working with the more general transport maps.
Covariance matrices arising from kernel functions are often fully
dense, but the approximate factors of their inverses can be sparse
if their ordering and sparsity pattern are chosen carefully.

\subsection{Ordering and sparsity pattern}
\label{subsec:ordering}

Although in this work we primarily focus on constructing
sparsity patterns, the chosen ordering critically affects
the accuracy of lower triangular sparsity patterns.
Vecchia originally proposed ordering points lexicographically, which is
most natural in a one-dimensional setting \cite{vecchia1988estimation}.
More recent work finds that in higher dimensions, exploiting
space-covering orderings leads to significantly better
approximation quality \cite{guinness2018permutation}.
Specifically, we use the (reverse of the) maximum-minimum (maximin)
ordering \cite{guinness2018permutation}, which has become popular for the
Vecchia approximation \cite{katzfuss2021general, Katzfuss2018, Cao2022} and
Cholesky factorization \cite{schafer2020compression, schafer2021sparse,
kang2021correlationbased, katzfuss2022scalable, Cao2023}\footnote{Some authors
instead and equivalently use the maximin ordering (without reversion), with
upper triangular (as opposed to lower triangular) Cholesky factors.}.
The reverse-maximin ordering \( i_1, \dotsc, i_N \) on a set of \( N \) points
\( \{ \vec{x}_i \}_{i = 1}^N \) is defined by first selecting the last index \(
i_N \) arbitrarily and then choosing for \( k = N - 1, N - 2, \dotsc, 1 \) the
index
\begin{align}
  \label{eq:maximin}
  i_k = \argmax_{i \in -\Order_{k + 1}} \; \min_{j \in \Order_{k + 1}}
    \norm{\vec{x}_i - \vec{x}_j}
\end{align}
where \( -\Order \defeq \{ 1, \dotsc, N \} \setminus \Order \)
and \( \Order_n \defeq \{ i_n, i_{n + 1}, \dotsc, i_N \} \),
i.e., select the point farthest from previously selected points.
We write \( i \prec j \) if \( i \) precedes \( j \) in the ordering and
define \( \ell_{i_k} \defeq \min_{j \in \Order_{k + 1}} \norm{\vec{x}_{i_k}
- \vec{x}_j} \), a length scale monotonically shrinking with decreasing
position in the ordering.
For robustness, we define the \( p \)-maximin ordering as replacing the
inner \( \min \) in \cref{eq:maximin} with the \( p \)-th smallest distance
(\( \infty \) if \( \card{\Order_{k + 1}} < p \)), generalizing the maximin
ordering (\( p = 1 \)).

Vecchia also originally proposed to select the sparsity set by Euclidean
distance \cite{vecchia1988estimation}, which, unlike the lexicographic
ordering, still remains widely used \cite{schafer2020compression,
schafer2021sparse, katzfuss2022scalable}.
Instead, we will select the sparsity pattern to directly
optimize the accuracy \( \Loss \) \cref{eq:generic_obj}.

\subsection{Review of KL-minimization}
\label{subsec:kl}

The Kullback-Leibler (KL) divergence between two probability distributions \( P
\) and \( Q \) is defined as \( \KL*{P}{Q} \defeq \E_P[\log(\frac{P}{Q})] \).
As the expected difference between true and approximate log-densities, the
KL divergence naturally judges the quality of an approximating distribution.
We view the positive-definite matrix \( \CM \in \Reals^{N \times N} \) as
the covariance matrix of a centered Gaussian \( \N(\vec{0}, \CM) \),
which we seek to approximate by a sparse approximate Cholesky factor \( L \in
\SpSet \) of its precision, \( \N(\vec{0}, (L L^{\top})^{-1}) \).
Like \cite{schafer2021sparse}, we measure error using the KL divergence,
specializing the generic optimization problem \cref{eq:generic_obj} to
\begin{align}
  \label{eq:L_obj}
  L \defeq \argmin_{\hat{L} \in \SpSet} \,
    \KL*{\N(\vec{0}, \CM)}
        {\N(\vec{0}, (\hat{L} \hat{L}^{\top})^{-1})}.
\end{align}
For centered multivariate Gaussians, the KL divergence
has a closed-form expression given by
\begin{align}
  \label{eq:kl}
  2 \KL*{\N(\vec{0}, \CM_1)}{\N(\vec{0}, \CM_2)} &=
    \trace(\CM_2^{-1} \CM_1) + \logdet(\CM_2) - \logdet(\CM_1) - N
\end{align}
where \( \CM_1, \CM_2 \in \Reals^{N \times N} \).
Using this expression for the KL divergence and optimizing for \( L \) yields
the following closed-form expression for the nonzero entries in the \( i \)-th
column of \( L \) with sparsity pattern \( s_i \), reproduced from Theorem 2.1
of \cite{schafer2021sparse}:
\begin{align}
  \label{eq:L_col}
  L_{s_i, i} &= \frac{\CM_{s_i, s_i}^{-1} \vec{e}_1}
    {\sqrt{\vec{e}_1^{\top} \CM_{s_i, s_i}^{-1} \vec{e}_1}}
\end{align}
where the notation for \( \CM_{s_i, s_i}^{-1} \) is from
\cref{eq:cond_cov_notation} and \( \vec{e}_1 \in \Reals^{\card{s_i}
\times 1} \) denotes the vector with first entry one and the rest zero.
We enforce the convention that \( i \) is the first entry
of \( s_i \), also implying that \( L \) is of full rank.
Plugging the optimal \( L \) \cref{eq:L_col} back into the KL
divergence \cref{eq:kl}, we obtain the objective as a function of
the sparsity pattern (see \cref{app:kl_L} for details; importantly,
the order of the KL divergence matters):
\begin{align}
  \label{eq:obj_chol}
  2 \KL*{\N(\vec{0}, \CM)}{\N(\vec{0}, (L L^{\top})^{-1})} &=
    \sum_{i = 1}^N
      \left [
        \log \left ( \CM_{i, i \mid s_i \setminus \{ i \}} \right ) -
        \log \left ( \CM_{i, i \mid i + 1:} \right )
      \right ].
\end{align}

This sum is the accumulated \emph{difference} in posterior log variance
for a series of independent regression problems: each to predict the \( i
\)-th variable given a subset of the variables after it in the ordering.
The term \( \log \left ( \CM_{i, i \mid s_i \setminus \{ i \}} \right )
\) obtained when restricted to the variables in the sparsity pattern \(
s_i \) is compared to the ground truth \( \log \left ( \CM_{i, i \mid i
+ 1:} \right ) \).
A similar decomposition of the KL divergence into independent
regression problems was observed in Equation (5) of
\cite{katzfuss2022scalable} for lower triangular transport maps.

Thus, picking the right sparsity pattern to minimize KL divergence reduces
to selecting the points \( s_i \) out of the possible candidates \( i + 1,
\dotsc, N \) that most reduce predictive error at point(s) of interest.
In the next section, we develop such a selection
algorithm for directed inference in Gaussian processes.
We apply this algorithm for sparsity selection of sparse Cholesky factors
in \cref{sec:chol_select} and extend it in \cref{sec:extensions}.

\section{Greedy selection for directed inference}
\label{sec:select}

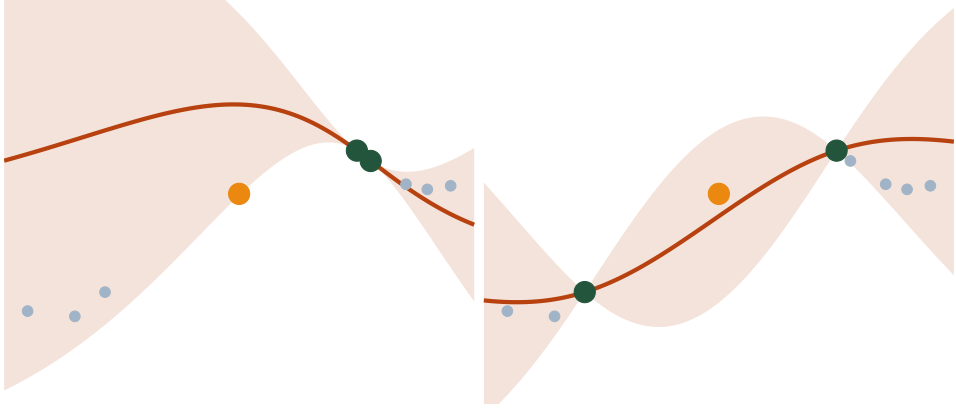
\begin{figure}[t]
  \centering
  \begin{tikzpicture}[baseline]
  \begin{axis}[
    width={0.4\linewidth},
    axis lines={none},
    scale only axis={true},
    xmin=-1, xmax=1, ymin=-1.5, ymax=1.75,
  ]
  \addplot [only marks, mark size=2, lightblue]
    table {figures/selection/data/train.csv};
  \addplot [only marks, mark size=4, orange]
    table {figures/selection/data/test.csv};
  \addplot [only marks, mark size=4, seagreen]
    table {figures/selection/data/knn_selected.csv};
  \addplot [ultra thick, rust]
    table {figures/selection/data/knn_mean.csv};
  \addplot [name path=upper, forget plot, draw=none, rust]
    table {figures/selection/data/knn_std_upper.csv};
  \addplot [name path=lower, forget plot, draw=none, rust]
    table {figures/selection/data/knn_std_lower.csv};
  \addplot [rust!15] fill between [of=upper and lower];
\end{axis}
\end{tikzpicture}%
  \begin{tikzpicture}[baseline]
  \begin{axis}[
    width={0.4\linewidth},
    axis lines={none},
    scale only axis={true},
    xmin=-1, xmax=1, ymin=-1.5, ymax=1.75,
  ]
  \addplot [only marks, mark size=2, lightblue]
    table {figures/selection/data/train.csv};
  \addplot [only marks, mark size=4, orange]
    table {figures/selection/data/test.csv};
  \addplot [only marks, mark size=4, seagreen]
    table {figures/selection/data/cknn_selected.csv};
  \addplot [ultra thick, rust]
    table {figures/selection/data/cknn_mean.csv};
  \addplot [name path=upper, forget plot, draw=none, rust]
    table {figures/selection/data/cknn_std_upper.csv};
  \addplot [name path=lower, forget plot, draw=none, rust]
    table {figures/selection/data/cknn_std_lower.csv};
  \addplot [rust!15] fill between [of=upper and lower];
\end{axis}
\end{tikzpicture}
  \caption{%
    Here, the \textcolor{lightblue}{blue} points are the
    \textcolor{lightblue}{candidates}, the \textcolor{orange}{orange}
    point is the \textcolor{orange}{target} point to predict
    at, and the \textcolor{seagreen}{green} points are the
    \textcolor{seagreen}{selected} points.
    The \textcolor{rust}{red} line is the \textcolor{rust}{conditional
    mean} \( \mu \), conditional on the selected points, and the \( \pm
    2 \sigma \) \textcolor{rust!60}{confidence interval} is shaded for
    the \textcolor{rust!60}{conditional variance} \( \var \).
    Each method has a budget of two points; the left panel shows selection
    by Euclidean distance and the right by conditional variance.
    Euclidean distance prefers the two points right of the target.
    However, a more balanced view of the situation is obtained when picking
    the slightly farther but
    more informative point to the left,
    reducing variance at the target and thereby reducing predictive error.
  }
  \label{fig:selection}
\end{figure}

In directed Gaussian process regression, we are given \( N \) points of
training data and predict at a target point with unknown value by selecting
the \( s \) points most ``informative'' to the target, \(s \ll N \).
From KL-minimization, the criterion for informativity
should be to minimize the variance of the target point,
conditional on the selected points \cref{eq:obj_chol}.
The variance objective was first described by \cite{cohn1996neural} for optimal
experimental design and later applied to directed Gaussian process inference
by \cite{gramacy2014local}, who refer to it as the active learning Cohn (ALC)
technique in honor of \cite{cohn1996neural}.
In addition, the variance objective is equivalent to maximizing the
\emph{mutual information} or \emph{information gain} with the target
point as well as to minimizing the expected mean squared error (see
\cref{app:mutual_info}).
The mutual information (in a slightly different context) is
also used by \cite{krause2008nearoptimal} for sensor placement.

In contrast to Euclidean distance \cite{vecchia1988estimation} or unconditional
correlation \cite{wada2013gaussian, kang2021correlationbased}, conditional
variance incentivizes the often contradictory demands of being near the
target point (nearby points have higher covariance), but away from previously
selected points (to avoid redundancy); the resulting spread-out selections
are illustrated in \cref{fig:selection}.
Even for isotropic kernels like the Mat{\'e}rn family, the
selections can be anisotropic, as shown in \cref{fig:cond_select}.

\subsection{A greedy approach}
\label{subsec:greedy_select}

Minimizing the conditional variance over all possible \(
\binom{N}{s} \) subsets is intractable, so we greedily select
the next point which most reduces the conditional variance.
Let \( \I = \{ i_j \}_{j = 1}^t \subseteq \Train \) be
the indices of previously selected training points.
For a newly selected index \( k \), we condition the current
covariance matrix on \( y_k \) according to the posterior
\cref{eq:quotient_rule}, resulting in the rank-one downdate
\begin{align}
  \label{eq:cond_select}
  \CM_{:, : \mid \I, k} &= \CM_{:, : \mid \I} - \vec{u} \vec{u}^{\top}, &
    \vec{u} &= \frac{\CM_{:, k \mid \I}}{\sqrt{\CM_{k, k \mid \I}}}.
\end{align}
The decrease in the variance of \( y_\Pred \) after
selecting \( k \) is given by \( u_\Pred^2 \), or
\begin{align}
  \label{eq:obj_gp}
  u_\Pred^2
  = \frac{\CM_{\Pred, k \mid \I}^2}{\CM_{k, k \mid \I}}
  = \frac{\Cov[y_\Pred, y_k \mid \I]^2}{\Var[y_k \mid \I]}
  = \Corr[y_\Pred, y_j \mid \I]^2 \Var[y_\Pred \mid \I].
\end{align}
To compute the objective \cref{eq:obj_gp} for each candidate index
\( j \), we start with the unconditional variance \( \CM_{j, j} \)
and covariance \( \CM_{\Pred, j} \), updating these quantities when
an index \( k \) is selected by Equation \cref{eq:cond_select}.
We have two efficient strategies to compute \( \vec{u} \) (as shown in
\cref{fig:alg_update}): either by maintaining the precision of selected
entries \( \CM_{\I, \I}^{-1} \) (\cref{alg:select_prec}) or by storing only
the \( \card{I} \) columns corresponding to selected points from the Cholesky
factor of the joint covariance matrix \( \CM \) (\cref{alg:select_chol});
both methods are detailed in \cref{app:select}.

\begin{figure}[th!]
  \centering
  \begin{minipage}[t]{0.49\textwidth}
    \begin{algorithm}[H]
      \caption{Point selection update \\ by explicit precision}
      \label{alg:update_prec}
      \input{figures/algorithms/update_prec.tex}
    \end{algorithm}
  \end{minipage}
  \hfill
  \begin{minipage}[t]{0.49\textwidth}
    \begin{algorithm}[H]
      \caption{Point selection update \\ by Cholesky factorization}
      \label{alg:update_chol}
      \input{figures/algorithms/update_chol.tex}
    \end{algorithm}
  \end{minipage}
  \caption{Algorithms for updates in single-target selection.}
  \label{fig:alg_update}
\end{figure}

To select \( s \) points out of \( N \) candidates, both approaches have
a time complexity of \( \BigO(N s^2) \) but differ in space complexity:
the precision takes \( \BigO(s^2) \) space, while the first \( s \)
columns of the Cholesky factor of \( \CM \) use \( \BigO(N s) \) space.
Both algorithms use \( \BigO(N) \) space
to store the conditional (co)variances.
The precision algorithm uses less memory than the Cholesky
algorithm, but the Cholesky algorithm is easier to implement and
roughly two times faster, which is why we use it in practice.

\section{Greedy selection for global approximation by KL-minimization}
\label{sec:chol_select}

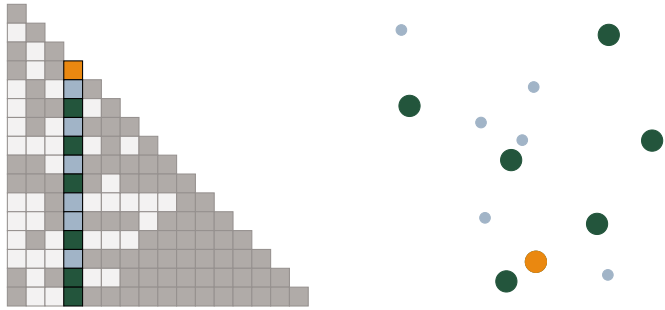
\begin{figure}[t]
  \centering
  \begin{tikzpicture}[scale=4/16]
  \filldraw[draw=nnzborder, fill=nnzcolor] (0, 0) rectangle (1, -1);
  \filldraw[draw=zeroborder, fill=zerocolor] (0, -1) rectangle (1, -2);
  \filldraw[draw=nnzborder, fill=nnzcolor] (0, -2) rectangle (1, -3);
  \filldraw[draw=nnzborder, fill=nnzcolor] (0, -3) rectangle (1, -4);
  \filldraw[draw=zeroborder, fill=zerocolor] (0, -4) rectangle (1, -5);
  \filldraw[draw=zeroborder, fill=zerocolor] (0, -5) rectangle (1, -6);
  \filldraw[draw=zeroborder, fill=zerocolor] (0, -6) rectangle (1, -7);
  \filldraw[draw=zeroborder, fill=zerocolor] (0, -7) rectangle (1, -8);
  \filldraw[draw=nnzborder, fill=nnzcolor] (0, -8) rectangle (1, -9);
  \filldraw[draw=nnzborder, fill=nnzcolor] (0, -9) rectangle (1, -10);
  \filldraw[draw=zeroborder, fill=zerocolor] (0, -10) rectangle (1, -11);
  \filldraw[draw=zeroborder, fill=zerocolor] (0, -11) rectangle (1, -12);
  \filldraw[draw=zeroborder, fill=zerocolor] (0, -12) rectangle (1, -13);
  \filldraw[draw=zeroborder, fill=zerocolor] (0, -13) rectangle (1, -14);
  \filldraw[draw=nnzborder, fill=nnzcolor] (0, -14) rectangle (1, -15);
  \filldraw[draw=nnzborder, fill=nnzcolor] (0, -15) rectangle (1, -16);
  \filldraw[draw=nnzborder, fill=nnzcolor] (1, -1) rectangle (2, -2);
  \filldraw[draw=zeroborder, fill=zerocolor] (1, -2) rectangle (2, -3);
  \filldraw[draw=zeroborder, fill=zerocolor] (1, -3) rectangle (2, -4);
  \filldraw[draw=nnzborder, fill=nnzcolor] (1, -4) rectangle (2, -5);
  \filldraw[draw=nnzborder, fill=nnzcolor] (1, -5) rectangle (2, -6);
  \filldraw[draw=nnzborder, fill=nnzcolor] (1, -6) rectangle (2, -7);
  \filldraw[draw=zeroborder, fill=zerocolor] (1, -7) rectangle (2, -8);
  \filldraw[draw=nnzborder, fill=nnzcolor] (1, -8) rectangle (2, -9);
  \filldraw[draw=nnzborder, fill=nnzcolor] (1, -9) rectangle (2, -10);
  \filldraw[draw=zeroborder, fill=zerocolor] (1, -10) rectangle (2, -11);
  \filldraw[draw=zeroborder, fill=zerocolor] (1, -11) rectangle (2, -12);
  \filldraw[draw=nnzborder, fill=nnzcolor] (1, -12) rectangle (2, -13);
  \filldraw[draw=zeroborder, fill=zerocolor] (1, -13) rectangle (2, -14);
  \filldraw[draw=zeroborder, fill=zerocolor] (1, -14) rectangle (2, -15);
  \filldraw[draw=zeroborder, fill=zerocolor] (1, -15) rectangle (2, -16);
  \filldraw[draw=nnzborder, fill=nnzcolor] (2, -2) rectangle (3, -3);
  \filldraw[draw=nnzborder, fill=nnzcolor] (2, -3) rectangle (3, -4);
  \filldraw[draw=zeroborder, fill=zerocolor] (2, -4) rectangle (3, -5);
  \filldraw[draw=nnzborder, fill=nnzcolor] (2, -5) rectangle (3, -6);
  \filldraw[draw=zeroborder, fill=zerocolor] (2, -6) rectangle (3, -7);
  \filldraw[draw=zeroborder, fill=zerocolor] (2, -7) rectangle (3, -8);
  \filldraw[draw=zeroborder, fill=zerocolor] (2, -8) rectangle (3, -9);
  \filldraw[draw=nnzborder, fill=nnzcolor] (2, -9) rectangle (3, -10);
  \filldraw[draw=nnzborder, fill=nnzcolor] (2, -10) rectangle (3, -11);
  \filldraw[draw=nnzborder, fill=nnzcolor] (2, -11) rectangle (3, -12);
  \filldraw[draw=zeroborder, fill=zerocolor] (2, -12) rectangle (3, -13);
  \filldraw[draw=zeroborder, fill=zerocolor] (2, -13) rectangle (3, -14);
  \filldraw[draw=nnzborder, fill=nnzcolor] (2, -14) rectangle (3, -15);
  \filldraw[draw=zeroborder, fill=zerocolor] (2, -15) rectangle (3, -16);
  \filldraw[draw=nnzborder, fill=nnzcolor] (4, -4) rectangle (5, -5);
  \filldraw[draw=zeroborder, fill=zerocolor] (4, -5) rectangle (5, -6);
  \filldraw[draw=nnzborder, fill=nnzcolor] (4, -6) rectangle (5, -7);
  \filldraw[draw=zeroborder, fill=zerocolor] (4, -7) rectangle (5, -8);
  \filldraw[draw=nnzborder, fill=nnzcolor] (4, -8) rectangle (5, -9);
  \filldraw[draw=nnzborder, fill=nnzcolor] (4, -9) rectangle (5, -10);
  \filldraw[draw=zeroborder, fill=zerocolor] (4, -10) rectangle (5, -11);
  \filldraw[draw=nnzborder, fill=nnzcolor] (4, -11) rectangle (5, -12);
  \filldraw[draw=zeroborder, fill=zerocolor] (4, -12) rectangle (5, -13);
  \filldraw[draw=nnzborder, fill=nnzcolor] (4, -13) rectangle (5, -14);
  \filldraw[draw=zeroborder, fill=zerocolor] (4, -14) rectangle (5, -15);
  \filldraw[draw=nnzborder, fill=nnzcolor] (4, -15) rectangle (5, -16);
  \filldraw[draw=nnzborder, fill=nnzcolor] (5, -5) rectangle (6, -6);
  \filldraw[draw=nnzborder, fill=nnzcolor] (5, -6) rectangle (6, -7);
  \filldraw[draw=nnzborder, fill=nnzcolor] (5, -7) rectangle (6, -8);
  \filldraw[draw=nnzborder, fill=nnzcolor] (5, -8) rectangle (6, -9);
  \filldraw[draw=zeroborder, fill=zerocolor] (5, -9) rectangle (6, -10);
  \filldraw[draw=zeroborder, fill=zerocolor] (5, -10) rectangle (6, -11);
  \filldraw[draw=nnzborder, fill=nnzcolor] (5, -11) rectangle (6, -12);
  \filldraw[draw=zeroborder, fill=zerocolor] (5, -12) rectangle (6, -13);
  \filldraw[draw=nnzborder, fill=nnzcolor] (5, -13) rectangle (6, -14);
  \filldraw[draw=zeroborder, fill=zerocolor] (5, -14) rectangle (6, -15);
  \filldraw[draw=nnzborder, fill=nnzcolor] (5, -15) rectangle (6, -16);
  \filldraw[draw=nnzborder, fill=nnzcolor] (6, -6) rectangle (7, -7);
  \filldraw[draw=zeroborder, fill=zerocolor] (6, -7) rectangle (7, -8);
  \filldraw[draw=nnzborder, fill=nnzcolor] (6, -8) rectangle (7, -9);
  \filldraw[draw=nnzborder, fill=nnzcolor] (6, -9) rectangle (7, -10);
  \filldraw[draw=zeroborder, fill=zerocolor] (6, -10) rectangle (7, -11);
  \filldraw[draw=nnzborder, fill=nnzcolor] (6, -11) rectangle (7, -12);
  \filldraw[draw=zeroborder, fill=zerocolor] (6, -12) rectangle (7, -13);
  \filldraw[draw=nnzborder, fill=nnzcolor] (6, -13) rectangle (7, -14);
  \filldraw[draw=nnzborder, fill=nnzcolor] (6, -14) rectangle (7, -15);
  \filldraw[draw=nnzborder, fill=nnzcolor] (6, -15) rectangle (7, -16);
  \filldraw[draw=nnzborder, fill=nnzcolor] (7, -7) rectangle (8, -8);
  \filldraw[draw=nnzborder, fill=nnzcolor] (7, -8) rectangle (8, -9);
  \filldraw[draw=nnzborder, fill=nnzcolor] (7, -9) rectangle (8, -10);
  \filldraw[draw=zeroborder, fill=zerocolor] (7, -10) rectangle (8, -11);
  \filldraw[draw=zeroborder, fill=zerocolor] (7, -11) rectangle (8, -12);
  \filldraw[draw=nnzborder, fill=nnzcolor] (7, -12) rectangle (8, -13);
  \filldraw[draw=nnzborder, fill=nnzcolor] (7, -13) rectangle (8, -14);
  \filldraw[draw=nnzborder, fill=nnzcolor] (7, -14) rectangle (8, -15);
  \filldraw[draw=nnzborder, fill=nnzcolor] (7, -15) rectangle (8, -16);
  \filldraw[draw=nnzborder, fill=nnzcolor] (8, -8) rectangle (9, -9);
  \filldraw[draw=nnzborder, fill=nnzcolor] (8, -9) rectangle (9, -10);
  \filldraw[draw=zeroborder, fill=zerocolor] (8, -10) rectangle (9, -11);
  \filldraw[draw=nnzborder, fill=nnzcolor] (8, -11) rectangle (9, -12);
  \filldraw[draw=nnzborder, fill=nnzcolor] (8, -12) rectangle (9, -13);
  \filldraw[draw=nnzborder, fill=nnzcolor] (8, -13) rectangle (9, -14);
  \filldraw[draw=nnzborder, fill=nnzcolor] (8, -14) rectangle (9, -15);
  \filldraw[draw=nnzborder, fill=nnzcolor] (8, -15) rectangle (9, -16);
  \filldraw[draw=nnzborder, fill=nnzcolor] (9, -9) rectangle (10, -10);
  \filldraw[draw=nnzborder, fill=nnzcolor] (9, -10) rectangle (10, -11);
  \filldraw[draw=nnzborder, fill=nnzcolor] (9, -11) rectangle (10, -12);
  \filldraw[draw=nnzborder, fill=nnzcolor] (9, -12) rectangle (10, -13);
  \filldraw[draw=nnzborder, fill=nnzcolor] (9, -13) rectangle (10, -14);
  \filldraw[draw=nnzborder, fill=nnzcolor] (9, -14) rectangle (10, -15);
  \filldraw[draw=nnzborder, fill=nnzcolor] (9, -15) rectangle (10, -16);
  \filldraw[draw=nnzborder, fill=nnzcolor] (10, -10) rectangle (11, -11);
  \filldraw[draw=nnzborder, fill=nnzcolor] (10, -11) rectangle (11, -12);
  \filldraw[draw=nnzborder, fill=nnzcolor] (10, -12) rectangle (11, -13);
  \filldraw[draw=nnzborder, fill=nnzcolor] (10, -13) rectangle (11, -14);
  \filldraw[draw=nnzborder, fill=nnzcolor] (10, -14) rectangle (11, -15);
  \filldraw[draw=nnzborder, fill=nnzcolor] (10, -15) rectangle (11, -16);
  \filldraw[draw=nnzborder, fill=nnzcolor] (11, -11) rectangle (12, -12);
  \filldraw[draw=nnzborder, fill=nnzcolor] (11, -12) rectangle (12, -13);
  \filldraw[draw=nnzborder, fill=nnzcolor] (11, -13) rectangle (12, -14);
  \filldraw[draw=nnzborder, fill=nnzcolor] (11, -14) rectangle (12, -15);
  \filldraw[draw=nnzborder, fill=nnzcolor] (11, -15) rectangle (12, -16);
  \filldraw[draw=nnzborder, fill=nnzcolor] (12, -12) rectangle (13, -13);
  \filldraw[draw=nnzborder, fill=nnzcolor] (12, -13) rectangle (13, -14);
  \filldraw[draw=nnzborder, fill=nnzcolor] (12, -14) rectangle (13, -15);
  \filldraw[draw=nnzborder, fill=nnzcolor] (12, -15) rectangle (13, -16);
  \filldraw[draw=nnzborder, fill=nnzcolor] (13, -13) rectangle (14, -14);
  \filldraw[draw=nnzborder, fill=nnzcolor] (13, -14) rectangle (14, -15);
  \filldraw[draw=nnzborder, fill=nnzcolor] (13, -15) rectangle (14, -16);
  \filldraw[draw=nnzborder, fill=nnzcolor] (14, -14) rectangle (15, -15);
  \filldraw[draw=nnzborder, fill=nnzcolor] (14, -15) rectangle (15, -16);
  \filldraw[draw=nnzborder, fill=nnzcolor] (15, -15) rectangle (16, -16);
  \filldraw[draw=colborder, fill=targetcolor] (3, -3) rectangle (4, -4);
  \filldraw[draw=colborder, fill=candcolor] (3, -4) rectangle (4, -5);
  \filldraw[draw=colborder, fill=selcolor] (3, -5) rectangle (4, -6);
  \filldraw[draw=colborder, fill=candcolor] (3, -6) rectangle (4, -7);
  \filldraw[draw=colborder, fill=selcolor] (3, -7) rectangle (4, -8);
  \filldraw[draw=colborder, fill=candcolor] (3, -8) rectangle (4, -9);
  \filldraw[draw=colborder, fill=selcolor] (3, -9) rectangle (4, -10);
  \filldraw[draw=colborder, fill=candcolor] (3, -10) rectangle (4, -11);
  \filldraw[draw=colborder, fill=candcolor] (3, -11) rectangle (4, -12);
  \filldraw[draw=colborder, fill=selcolor] (3, -12) rectangle (4, -13);
  \filldraw[draw=colborder, fill=candcolor] (3, -13) rectangle (4, -14);
  \filldraw[draw=colborder, fill=selcolor] (3, -14) rectangle (4, -15);
  \filldraw[draw=colborder, fill=selcolor] (3, -15) rectangle (4, -16);
\end{tikzpicture}%
  \qquad
  \begin{tikzpicture}[baseline]
  \begin{axis}[
    width={4cm},
    height={4cm},
    axis lines={none},
    scale only axis=true,
  ]
  \draw [white, line width=0] (-0.1, -0.1) -- (-0.1,  1.1);
  \draw [white, line width=0] ( 1.1, -0.1) -- ( 1.1,  1.1);
  \draw [white, line width=0] (-0.1, -0.1) -- (-1.1, -0.1);
  \draw [white, line width=0] (-0.1,  1.1) -- (-1.1,  1.1);
  \addplot [only marks, mark size=2, lightblue] table
    {figures/points/data/candidates.csv};
  \addplot [only marks, mark size=4, seagreen]  table
    {figures/points/data/selected.csv};
  \addplot [only marks, mark size=4, orange]    table
    {figures/points/data/target.csv};
  \end{axis}
\end{tikzpicture}
  \caption{%
    For a column of a Cholesky factor in isolation, the
    \textcolor{orange}{target} point is the \textcolor{orange}{diagonal}
    entry, \textcolor{lightblue}{candidates} are \textcolor{lightblue}{below}
    it, and the \textcolor{seagreen}{selected} entries are added to
    the \textcolor{seagreen}{sparsity pattern}.
    Points violating lower triangularity are
    not shown.
    Thus, sparsity selection in Cholesky factorization
    (left panel) is analogous to training point selection
    in directed Gaussian process regression (right panel).
  }
  \label{fig:select_chol}
\end{figure}

Directed Gaussian process regression infers the
\emph{local} distribution at points of interest.
We now turn our attention to \emph{global} approximation of the entire Gaussian
process; given a kernel function \( \K(\vec{x}, \vec{x}') \) and a set of \(
N \) points \( \{ \vec{x}_i \}_{i = 1}^N \), we have the covariance matrix \(
\CM_{i, j} = K(\vec{x}_i, \vec{x}_j) \), for which we seek a sparse approximate
Cholesky factor \( L \) of the precision, \( L L^{\top} \approx \CM^{-1} \).
We first order the points by the reverse-maximin ordering described in
\cref{subsec:ordering} and then apply the selection
algorithm developed in \cref{sec:select} to form the sparsity pattern.
For the \( i \)-th column of \( L \), the target point is the \( i \)-th point
in the ordering, the candidate points are those satisfying lower triangularity
(after the target in the ordering), and running the selection algorithm for
the desired number of nonzeros entries picks out indices which we add to the
sparsity set \( s_i \); this process is illustrated in \cref{fig:select_chol}.
Finally, we compute the values of the selected nonzero
indices by the closed-form expression \cref{eq:L_col}.
Both sparsity selection and computing entries are
embarrassingly parallel over \( L \)'s columns.

Using the single-target algorithm from \cref{subsec:greedy_select}
has time complexity \( \BigO(N C s^2) \) to select \( s \)
nonzero entries out of \( C \) candidates for \( N \) columns.
Computing the corresponding values \( \CM_{s_i, s_i}^{-1}
\vec{e}_1 \) has time complexity \( \BigO(N s^3) \).
Sparsity selection has the same complexity as entry computation
if the number of candidates \( C \) is \( \BigO(s) \), suggesting
the need to limit the number of candidates considered.
In practice, we pick the candidate set to be the nearest neighbors
of the point of interest as \cite{gramacy2014local} does;
specifically, we use the framework of \cite{schafer2021sparse}
which considers all points within a radius proportional to the
length scale from the reverse-maximin ordering \cref{eq:maximin}.
The improved accuracy of our method for a given number of nonzero entries
observed in \cref{sec:experiments} allows choosing a smaller \( s \) than
when using the non-adaptive method of \cite{schafer2021sparse}, thus enabling
computational savings.
When solving sequences of related problems, re-using the
adaptively selected sparsity set allows additional savings.
Other possibilities for selecting candidates include replacing the Euclidean
distance with a correlation distance \cite{kang2021correlationbased},
additionally including the last points in the ordering to exploit spectral
decay of the covariance matrix as in the Nystr{\"o}m approximation, and
randomly sampling points.
We defer a detailed comparison of different candidate sets to future work.

\section{Extensions}
\label{sec:extensions}

We now consider extensions of the single-column method
to aggregated (or supernodal) Cholesky factorization
and determining the number of nonzeros per column.

\subsection{Aggregated sparsity pattern}
\label{subsec:aggregated}

We now derive a similar decomposition of the KL divergence
if the same sparsity pattern is reused for multiple columns,
known as aggregated or supernodal Cholesky factorization.
Aggregation can lead to substantial time
and space savings \cite{schafer2021sparse}.
For this section, we focus on a single group \( \tilde{i}
= \{i_1, \dotsc, i_m \} \) obtained by aggregating the
column indices \( i_1 \succ i_2 \succ \dotsb \succ i_m \).
Let \( \tilde{i} \) have aggregated selected entries \(
s_{\tilde{i}} \) satisfying \( s_{\tilde{i}} \supseteq \tilde{i}
\) to guarantee that the Cholesky factor has full rank.
Let \( \tilde{s} \defeq s_{\tilde{i}} \setminus \tilde{i} \)
be the selected entries excluding the columns in the group.
The sparsity pattern for the \( k \)-th column in the group is then the
aggregated selected entries excluding the entries that violate lower
triangularity, \( s_k \defeq \{ j \in s_{\tilde{i}} : j \succeq k \} \).
Assuming every entry of \( \tilde{s} \) is after every index in \( \tilde{i}
\), then \( s_k = \tilde{s} \cup \{ j \in \tilde{i} : j \succeq k \} \).
This condition is guaranteed if the aggregated columns are adjacent in the
ordering, for example; we defer handling the general case to the next section.
The KL divergence \cref{eq:obj_chol} restricted to
the contribution from the group \( \tilde{i} \) is
\begin{align}
  \label{eq:obj_mult}
  \sum_{i \in \tilde{i}}
    \log \left (\CM_{i, i \mid s_i \setminus \{ i \} } \right ) &=
    \logdet(\CM_{\tilde{i}, \tilde{i} \mid \tilde{s}})
\end{align}
from \cref{app:kl_agg}.
Thus, the generalization of the posterior variance
\cref{eq:obj_chol} to aggregated columns is their log
determinant conditional on (well-behaved) selected entries.
We briefly discuss what happens when
selected entries are \emph{between} columns.

\subsubsection{Nonadjacent or partial aggregation}
\label{subsubsec:partial}

Let the random variables corresponding to the indices \( \tilde{i} = \{
i_1, \dotsc, i_m \} \) be collected in a vector \( \vec{y} = [y_1, \dotsc,
y_m]^{\top} \) with joint density \( \vec{y} \sim \N(\vec{0}, \CM) \).
We select a random variable with index \( k \) and define \emph{partial}
conditioning to mean that \( k \) conditions all but the first \( p \)
variables (recall that the indices are sorted w.r.t \( \succ \), so if
\( k \) conditions a variable, it conditions all those afterwards).
We denote the partial conditioning of \( y \) as \( \vec{y}_{\parallel
k} \defeq [y_1, \dotsc, y_p, y_{p + 1 \mid k}, \dotsc, y_{m \mid k}]^{\top}
\) and its covariance matrix \( \Cov[\vec{y}_{\parallel k}] \) as
\begin{align}
  \label{eq:chol_partial}
  \CM_{\tilde{i}, \tilde{i} \parallel k} &\defeq
  \Cov[\vec{y}_{\parallel k}] =
  \begin{pmatrix}
    L_{:p} L_{:p}^{\top} &
    L_{:p} {L'}_{p + 1:}^{\top} \\
    {L'}_{p + 1:} L_{:p}^{\top} &
    {L'}_{p + 1:} {L'}_{p + 1:}^{\top}
  \end{pmatrix} =
  \begin{pmatrix}
    L_{:p} \\
    {L'}_{p + 1:}
  \end{pmatrix}
  \begin{pmatrix}
    L_{:p} \\
    {L'}_{p + 1:}
  \end{pmatrix}^{\top}
\end{align}
where \( L = \chol(\CM) \) and \( L' = \chol(\CM_{:, : \mid k}) \).
See \cref{fig:partial_factor} for an
illustration and \cref{app:partial} for details.
Armed with this representation, we compute \(
\logdet(\CM_{\tilde{i}, \tilde{i} \parallel k}) \):
\begin{align}
  \label{eq:partial_kl}
  \sum_{i \in \tilde{i}}
    \log \left (\CM_{i, i \mid s_i \setminus \{ i \} } \right ) &=
    \logdet(\CM_{\tilde{i}, \tilde{i} \parallel k}).
\end{align}
Like the aggregated case \cref{eq:obj_mult}, minimizing the log determinant
of the \emph{partially} conditioned covariance matrix \cref{eq:chol_partial}
is the same as minimizing the KL divergence \cref{eq:obj_chol}.

\subsection{Supernodes and blocked selection}
\label{subsec:mult_select}

For multiple (\( m > 1 \)) target points, \cite{gramacy2014local} suggests
independently applying the single-target algorithm to each target.
Instead, we will use \emph{the same} selected points for \emph{all} the
target points, essentially speeding up selection by a factor of \( m \).
Furthermore, the cost of computing the entries of the
resulting Cholesky factor is reduced by this aggregation.
The primary downside is reduced accuracy per sparsity entry
since each target no longer receives individual attention.
We mitigate this by paying heed to the ``two birds with
one stone'' maxim, or by considering a candidate's
simultaneous effect on \emph{all} prediction points.
In practice, this approach yields better accuracy
per unit of time than single-target selection.

The first question is how to generalize the objective for
a single target \cref{eq:obj_gp} to multiple targets.
Continuing with KL-minimization \cref{eq:obj_mult}, the criterion
should be to minimize the log determinant of the posterior
covariance matrix, \( \logdet(\CM_{\Pred, \Pred \mid I}) \).
This objective, known as D-optimal design in the literature
\cite{krause2008nearoptimal}, can be intuitively interpreted as a volume of
uncertainty or as a scaling factor in the density of multivariate Gaussians.
In addition, it is equivalent to maximizing mutual information
since the differential entropy of a Gaussian strictly increases
with its log determinant (see \cref{app:mutual_info}).

We want to quickly compute how selecting an
index \( k \) affects the log determinant.
By application of the matrix determinant lemma
(the details are in \cref{app:logdet_downdate}),
\begin{align}
  \label{eq:greedy_mult}
  \logdet(\CM_{\Pred, \Pred \mid \I, k})  - \logdet(\CM_{\Pred, \Pred \mid \I})
  &= \log(\CM_{    k,     k \mid \I, \Pred}) - \log(\CM_{    k,     k \mid \I}).
\end{align}
Equation \cref{eq:greedy_mult} swaps the roles of the targets
and the candidate: the \emph{candidate} is now conditioned
by the \emph{targets}, reducing to single-target selection.
Using the recipes from \cref{subsec:greedy_select} to compute conditional
variances, we can compute the objective by maintaining a data structure
for each term: one for \( \CM_{k, k \mid \I, \Pred} \) and the other for
\( \CM_{k, k \mid \I} \).
By the quotient rule \( \CM_{k, k \mid \I, \Pred} = \CM_{k,
k \mid \Pred, \I} \), so we can condition on the prediction
points \emph{before} any points have been selected.
After this initialization, we repeatedly update both data structures after
selecting the best candidate by the objective \cref{eq:greedy_mult}.

We have two strategies from the two approaches of the single-target
algorithm: one maintaining the precision of the selected entries \( \CM_{\I,
\I}^{-1} \) as well as of the target points \( \CM_{\Pred, \Pred}^{-1} \)
(\cref{alg:select_mult_prec}) and the other simply storing two Cholesky factors
of the joint covariance matrix \( \CM \) (\cref{alg:select_mult_chol}); both
methods are detailed in \cref{app:mult_select}.

Both approaches have a time complexity of \( \BigO(N s^2 + N m^2 + m^3) \) to
select \( s \) points out of \( N \) candidates for \( m \) targets, again
differing in space complexity: although using more memory, the Cholesky
approach is preferred for simplicity and performance.

\paragraph{Partial selection}

The multiple-target algorithm implicitly assumes that a candidate
conditions \emph{every} target point. 
However, candidates
can also condition only a subset of the targets in the aggregated
Cholesky factorization setting of \cref{subsubsec:partial}.
Proper ``partial'' selection accounting for this matches the asymptotic time
and space complexities of the multiple-target algorithm by storing a partial
Cholesky factor. Details are provided in \cref{subsec:partial_select} and
\cref{alg:select_partial}.

\subsection{Aggregated Cholesky}

In an aggregated sparsity pattern, columns are partitioned into groups and
selecting an index for a group \( \tilde{i} \) adds it to the sparsity.

We group columns by the framework of \cite{schafer2021sparse}, which aggregates
points that are close both geometrically as well as in the ordering.
To select sparsity entries, the targets are all points in \( \tilde{i} \)
and the candidates are the union of the nearest neighbors to each target.
If every candidate \( k \) satisfies \( k \succ \max{\tilde{i}} \), which
occurs if the group is contiguous in the ordering; then every candidate
conditions every target and so the multiple-target selection algorithm
(\cref{subsec:mult_select}) can be directly applied.
However, we empirically observe that forcing this condition irreparably
damages the accuracy of the resulting factor: forming groups contiguous in
the ordering no longer guarantees that grouped points are spatially close,
and removing candidates between targets filters many of them out.
If the condition is not forced, then selecting candidates can
condition subsets of the group; the multiple-target algorithm
now systematically overestimates the effect on targets.
Using the partial selection algorithm (\cref{subsec:partial_select})
instead on unmodified grouping and candidate sets
significantly improves the approximation quality.

Because the sparsity patterns for columns in the same group are subsets
of each other, we can efficiently compute the group's entries in \( L \)
\cref{eq:L_col} together in the time complexity for a single column (see
\cref{app:L_mult} or Algorithm 3.2 of \cite{schafer2021sparse}).
If each group has \( m \) points, both the multiple-target and
partial algorithms have time complexity \( \BigO(C s^2 + C m^2
+ m^3) \) to select \( s \) points out of \( C \) candidates.
Over \( N/m \) groups the time complexity for both selection and entry
computation is \( \BigO(\frac{N}{m} (C s^2 + C m^2 + m^3 + s^3)) \),
simplifying to \( \BigO(\frac{N C s^2}{m}) \) assuming \( m = \BigO(s) \),
a \( m \) times improvement over non-aggregated factorization.
For \(m \approx s\), this complexity further simplifies to \( \BigO(N C s) \).
Better time complexity yields denser and thereby
more accurate factors in the same amount of time.
However, the sparsity pattern is no longer tailored to particular columns
since it is shared within a group.
This means the aggregated factor is less efficient
at reducing the KL divergence per nonzero.

\subsection{Allocating nonzeros by global greedy selection}
\label{subsec:global_greedy}

Given a budget on the total number of nonzeros, one
must decide how many nonzeros to assign to each column.
We recommend the simple strategy of distributing nonzeros as evenly as
possible, maximizing computational efficiency since denser columns have
an outsized impact on the computational time from the cubic scaling cost
with the number of nonzeros.

In inhomogeneous geometries, where certain points benefit from more
nonzeros than others, a principled way of distributing nonzeros is
to minimize KL divergence end-to-end, as we proposed for sparsity
selection in \cref{sec:select}.
The \emph{local} greedy algorithms select the sparsity entry that
minimizes prediction error at \emph{particular} columns of interest.
In \emph{global} greedy selection, we pick from \emph{any} column the
candidate that minimizes the overall KL divergence \cref{eq:obj_chol}.
We maintain a priority queue containing all candidates from every
column, keyed by the candidate's effect on the KL divergence.
The data structure must support removing the largest element from the
queue, as well as updating the value for an element in the queue.
Both operations have time complexity \( \BigO(\log n) \) for \( n \)
elements if implemented as an array-backed binary heap, for example.

The greedy selection algorithms already compute the effect of
an entry on the KL divergence, up to monotonically increasing
transformations (which preserve the ranking of candidates).
But in the global context, if different columns use different
transformations, then the ranking of candidates between columns is skewed.
We describe the necessary modifications to
compute exactly the difference in KL divergence.

\subsubsection{Single column selection}
\label{subsubsec:single_column}

Selecting an entry \( k \) for a single target only affects its conditional
variance, so exactly one term in the KL divergence \cref{eq:obj_chol} changes,
\begin{align}
  \argmin_k \left [
    \log(\CM_{\Pred, \Pred \mid \I, k}) - \log(\CM_{\Pred, \Pred \mid \I})
  \right ] &=
    \argmin_k \left (
      \CM_{\Pred, \Pred \mid \I, k}
    \right ) \CM_{\Pred, \Pred \mid \I}^{-1}.
  \shortintertext{
    Using the original objective \cref{eq:obj_gp} to
    compute the change in variance from selecting \( k \),
  }
  \label{eq:global_obj}
  \argmin_k \left (
    \CM_{\Pred, \Pred \mid \I} -
      \frac{\CM_{\Pred, k \mid \I}^2}{\CM_{k, k \mid \I}}
  \right ) \CM_{\Pred, \Pred \mid \I}^{-1} &=
    \argmax_k \; \Corr[y_\Pred, y_k \mid \I]^2
\end{align}
where the new objective \cref{eq:global_obj} is easily computed as
the original objective \cref{eq:obj_gp} divided by the target's
conditional variance, the percentage the decrease in variance takes up.

\subsubsection{Aggregated selection}
\label{subsubsec:mult_column}

The multiple-target algorithm already computes the exact
difference in log determinant after selecting a candidate.
The partial selection algorithm computes the log determinant itself, not
the difference, so the log determinant before the selection needs to be
subtracted, easily computed as the sum of the squared ``diagonal'' entries
of \( L \) corresponding to target points from \cref{eq:partial_diag}.

One improvement is to consider that the number of nonzeros added
from selecting a candidate is the number of targets conditioned.
In practice, it is often better to measure the
average decrease in KL divergence per sparsity entry.
For the multiple-target algorithm, this modification makes no difference
within a group since every candidate conditions every target, but can
make a difference in the global setting if groups have different sizes.
This modification can make a difference both within a column
and globally for the partial algorithm since candidates within
the same group can condition a different number of targets.

\section{Numerical experiments}
\label{sec:experiments}

All experiments ran on the Partnership for an Advanced Computing Environment
(PACE) Phoenix cluster at the Georgia Institute of Technology, with 8
cores of a Intel Xeon Gold 6226 CPU @ 2.70GHz and 22 GB of RAM per core.
The code is written in Python using standard scientific libraries
\texttt{numpy} \cite{harris2020array}, \texttt{scipy} \cite{virtanen2020scipy},
\texttt{scikit-learn} \cite{pedregosa2011scikitlearn}, \texttt{matplotlib}
\cite{hunter2007matplotlib} as well as Cython \cite{behnel2011cython} which
provides direct transpilation of Python code into C.
Cython also allows Python code to access native C interfaces
to the \texttt{BLAS} and Intel \texttt{oneMKL} libraries.
Code for all numerical experiments can be found at
\href{https://github.com/stephen-huan/conditional-knn}
{https://github.com/stephen-huan/conditional-knn}.

\pgfplotsset{
  cycle list={
    {very thick, lightblue, style=dashed,         mark=*},
    {very thick, seagreen,  style=dashdotted,     mark=o},
    {very thick, silver,    style=densely dotted, mark=triangle*},
    {very thick, orange,    style=solid,          mark=square*},
    {very thick, rust,      style=dotted,         mark=square},
    {very thick, joshua,    style=loosely dashed, mark=diamond*},
  }
}

\begin{figure}[t]
  \centering
  \begin{tikzpicture}[baseline, scale=0.75]
  \begin{axis}[
    grid={major},
    xlabel={\( N \)},
    ylabel={KL divergence},
    legend entries={%
      \( \rho \)-ball, \( \rho \)-ball (agg.),
      \( k \)-NN, select, select (agg.)
    },
    legend pos={north west},
  ]
  \addplot table {figures/cholesky/data/n_kl_div_KL.csv};
  \addplot table {figures/cholesky/data/n_kl_div_KL_agg.csv};
  \addplot table {figures/cholesky/data/n_kl_div_select-KNN.csv};
  \addplot table {figures/cholesky/data/n_kl_div_select.csv};
  \addplot table {figures/cholesky/data/n_kl_div_select_agg.csv};
  \end{axis}
\end{tikzpicture}%
  \begin{tikzpicture}[baseline, scale=0.75]
  \begin{axis}[
    grid={major},
    xlabel={\( N \)},
    ylabel={Time (seconds)},
  ]
  \addplot table {figures/cholesky/data/n_time_KL.csv};
  \addplot table {figures/cholesky/data/n_time_KL_agg.csv};
  \addplot table {figures/cholesky/data/n_time_select-KNN.csv};
  \addplot table {figures/cholesky/data/n_time_select.csv};
  \addplot table {figures/cholesky/data/n_time_select_agg.csv};
  \end{axis}
\end{tikzpicture}
  \caption{%
    Accuracy (left) and computational time (right) of
    Cholesky factorization methods with varying number
    of points \( N \) and fixed density \( \rho = 2 \).
    ``\( \rho \)-ball'' is the baseline from \cite{schafer2021sparse},
    ``\( k \)-NN'' is selection by \( k \)-nearest neighbors, ``select''
    is conditional selection, and ``(agg.)'' denotes aggregation.
  }
  \label{fig:chol_n}
\end{figure}
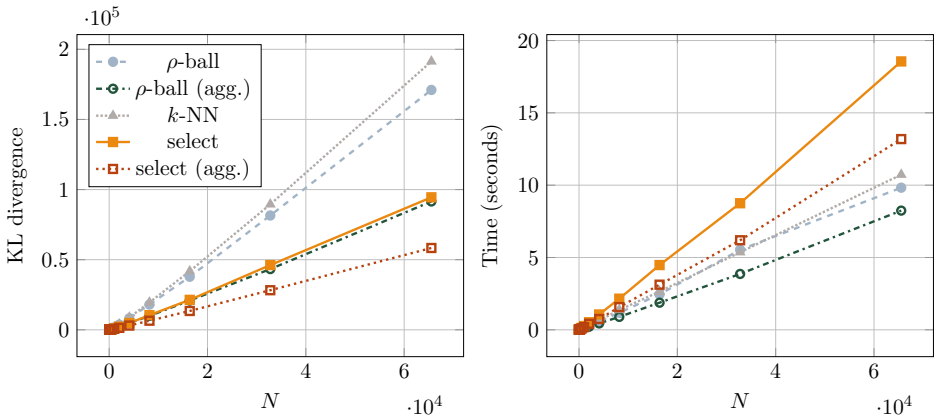

\subsection{Cholesky factorization}
\label{subsec:chol_exp}

We empirically verify that conditional sparsity selection
produces more accurate sparse Cholesky factors than
selection by Euclidean distance at the same density.
We take \( N \) points on a slightly perturbed regular
grid in \( [0, 1]^2 \) and use a Mat{\'e}rn kernel with
smoothness \( \nu = 5/2 \) and length scale \( \ell = 1 \).

As a baseline for comparison, we use the single-column and aggregated variants
of the KL-minimization framework of \cite{schafer2021sparse}, which orders
points by the reverse-maximin ordering described in \cref{subsec:ordering}
and forms the sparsity pattern by selecting all points within a radius of
\( \rho \ell_i \) to the \( i \)-th point, where \( \rho \geq 1 \) is a
tuning parameter for density and \( \ell_i \) is the length scale from the
reverse-maximin ordering.
We also try selecting points by \( k \)-nearest neighbors (\( k \)-NN),
where \( k \) is chosen to match the number of nonzeros of the baseline.

For our method, we run the baseline with a larger \( \rho' = \rho_s
\cdot \rho \) to get an initial candidate set where \( \rho_s \)
is a tuning parameter for the number of candidates considered.
If not stated otherwise, \( \rho_s = 2 \) is used in the following experiments.
We then use the single-column and partial variants of the conditional
selection algorithm described in \cref{sec:select} to select the actual
sparsity entries from the candidate set; in these experiments, each
column receives \( k \) nonzeros where \( k \) is chosen to match the
original density \( \rho \) of the baseline factor.
In this setting, we found that using the global selection procedure
described in \cref{subsec:global_greedy} to determine the number of
nonzeros for each column led to little improvement in accuracy at a
significant performance penalty.

We aggregate columns by the procedure of \cite{schafer2021sparse}: pick
the first (w.r.t \( \prec \)) index \( i \) that has not been aggregated
and create the group \( \{ j: j \in s_i, \ell_j \leq \lambda \ell_i \} \)
where \( \lambda \geq 1 \) is a tuning parameter for group size; repeat
until all indices have been aggregated.
We generate the groups using the sparsity pattern
from the baseline factor and use the same groups for
aggregated conditional selection for a fair comparison.
In all experiments we use \(\lambda = 1.5 \)
as recommended by \cite{schafer2021sparse}.

As \cref{fig:chol_n} shows, the KL divergence and computational
time increase linearly with the number of points for all methods.
Conditional methods are more accurate than their unconditional
counterparts and the denser aggregated variants are both more
accurate and faster than their non-aggregated counterparts.

\begin{figure}[t]
  \centering
  \begin{tikzpicture}[baseline, scale=0.75]
  \begin{semilogyaxis}[
    grid={major},
    xlabel={Density (\( \mathsf{nnz} / N^2 \))},
    ylabel={KL divergence},
    legend entries={%
      \( \rho \)-ball, \( \rho \)-ball (agg.),
      \( k \)-NN, select, select (agg.)
    },
    legend pos={north east},
  ]
  \addplot table {figures/cholesky/data/rho_kl_div_KL.csv};
  \addplot table {figures/cholesky/data/rho_kl_div_KL_agg.csv};
  \addplot table {figures/cholesky/data/rho_kl_div_select-KNN.csv};
  \addplot table {figures/cholesky/data/rho_kl_div_select.csv};
  \addplot table {figures/cholesky/data/rho_kl_div_select_agg.csv};
  \end{semilogyaxis}
\end{tikzpicture}%
  \begin{tikzpicture}[baseline, scale=0.75]
  \begin{loglogaxis}[
    grid={major},
    xlabel={Time (seconds)},
  ]
  \addplot table {figures/cholesky/data/rho_time_KL_kl_div.csv};
  \addplot table {figures/cholesky/data/rho_time_KL_agg_kl_div.csv};
  \addplot table {figures/cholesky/data/rho_time_select-KNN_kl_div.csv};
  \addplot table {figures/cholesky/data/rho_time_select_kl_div.csv};
  \addplot table {figures/cholesky/data/rho_time_select_agg_kl_div.csv};
  \end{loglogaxis}
\end{tikzpicture}
  \caption{
    The left panel shows the KL divergence with varying
    density \( \rho \) and the right panel shows the accuracy
    to computational time trade-off over varying \( \rho \).
    The number of points is \( N = 2^{16} \).
    The regular grid geometry limits the possible improvements due
    to the selection, but we still observe a slight improvement.
  }
  \label{fig:chol_rho}
\end{figure}
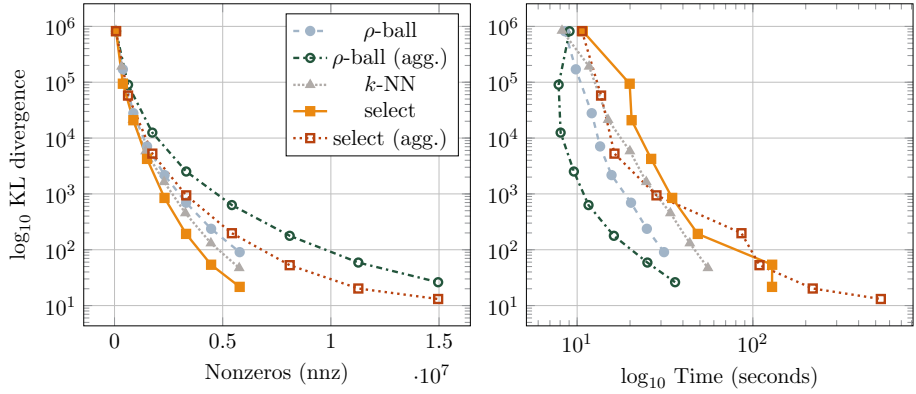

\cref{fig:chol_rho} shows that the conditional selection methods achieve
significantly better KL divergence compared to their unconditional
counterparts for the same number of entries of the sparsity set.
However, they do not achieve better accuracy per
unit time cost from the increased cost of selection.
Aggregation results in better accuracy per unit time cost but worse accuracy
per nonzero entry, which may impact their computational efficiency in
downstream tasks that depend on factor density, such as preconditioning.
Under the assumptions for rigorous error estimates in
\cite{schafer2021sparse}, its authors show that the asymptotic convergence
rates are the same when using KL, operator norm, or Frobenius norm.
For the sake of completeness, we report the latter two
measures of error in \cref{fig:chol_n_norm,fig:chol_rho_norm}.

\subsection{Gaussian process regression}
\label{subsec:gp_exp}

For Gaussian process regression we use the ``predictions points first''
method of \cite{schafer2021sparse}, which computes a sparse Cholesky factor
of the joint covariance matrix between the training and prediction points,
where prediction points are ordered before training points.
The desired posterior mean and covariance can then be
computed efficiently from sparse submatrices of the factor.
See subsection 4.2.1, Appendix D.1, or Algorithm D.1
of \cite{schafer2021sparse} for additional details.

For both datasets we use a \( 2 \)-maximin ordering (\cref{subsec:ordering}).

\begin{figure}[t]
  \centering
  \includegraphics{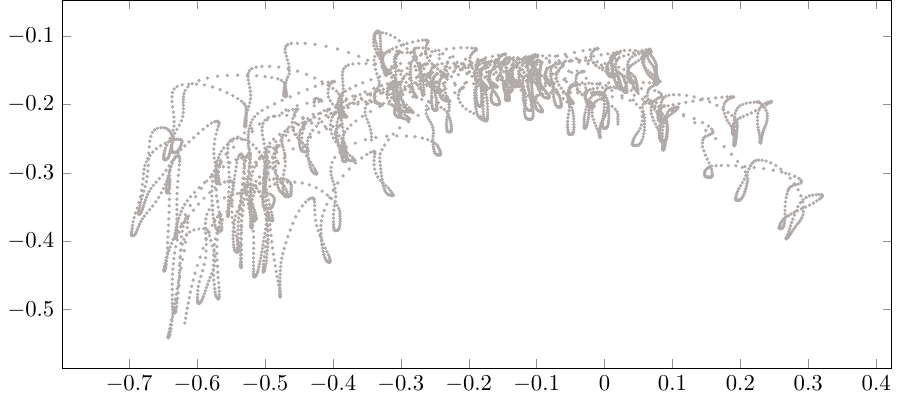}
  \caption{%
    The first 5,000 points of the SARCOS training
    data, visualizing the first two features.
  }
  \label{fig:sarcos}
\end{figure}

\begin{figure}[t]
  \centering
  \begin{tikzpicture}[baseline, scale=0.75]
  \begin{loglogaxis}[
    grid={major},
    xlabel={Density (\( \mathsf{nnz} / N^2 \))},
    ylabel={RMSE},
  ]
  \addplot table {figures/gp/sarcos/data/rho_loss_KL.csv};
  \addplot table {figures/gp/sarcos/data/rho_loss_KL_agg.csv};
  \addplot table {figures/gp/sarcos/data/rho_loss_select-KNN.csv};
  \addplot table {figures/gp/sarcos/data/rho_loss_select.csv};
  \addplot table {figures/gp/sarcos/data/rho_loss_select_agg.csv};
  \end{loglogaxis}
\end{tikzpicture}%
  \begin{tikzpicture}[baseline, scale=0.75]
  \begin{loglogaxis}[
    grid={major},
    xlabel={Time (seconds)},
    legend entries={%
      \( \rho \)-ball, \( \rho \)-ball (agg.),
      \( k \)-NN, select, select (agg.)
    },
    legend pos={north east},
    legend style={nodes={scale=0.9, transform shape}},
  ]
  \addplot table {figures/gp/sarcos/data/rho_time_KL_loss.csv};
  \addplot table {figures/gp/sarcos/data/rho_time_KL_agg_loss.csv};
  \addplot table {figures/gp/sarcos/data/rho_time_select-KNN_loss.csv};
  \addplot table {figures/gp/sarcos/data/rho_time_select_loss.csv};
  \addplot table {figures/gp/sarcos/data/rho_time_select_agg_loss.csv};
  \end{loglogaxis}
\end{tikzpicture}
  \caption{%
    We perform Gaussian process regression on the (modified)
    SARCOS dataset (as visualized in \cref{fig:sarcos}) by sparse
    Cholesky factorization of the joint covariance matrix.
    The left panel shows the difference in RMSE from exact Gaussian process
    regression using the same training points with varying density \( \rho \).
    The right panel shows the accuracy to computational
    time trade-off over varying \( \rho \).
  }
  \label{fig:gp_sarcos_rho}
\end{figure}
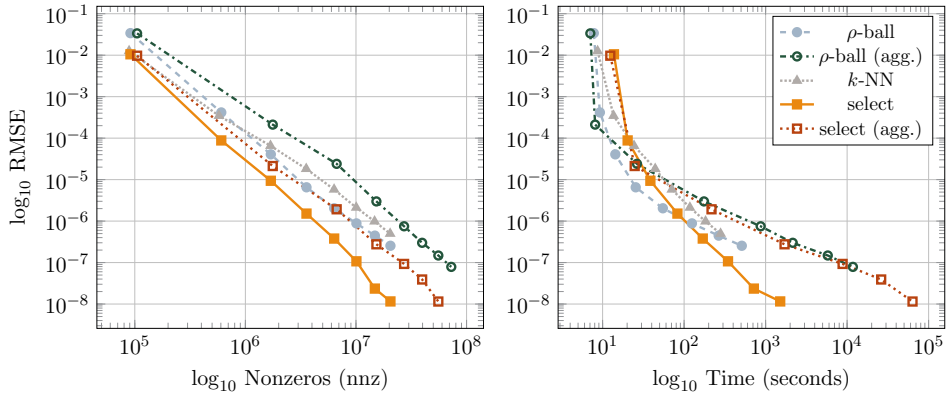

\subsubsection{SARCOS dataset}

We use the SARCOS dataset \cite{vijayakumar2000locally} generated from
an inverse dynamics problem on a 7 degrees-of-freedom robotic arm.
Given 7 joint positions, velocities, and accelerations for 21 features
in total, the goal is to infer the torque of just the first joint.
The dataset is available online at \url{https://gaussianprocess.org/gpml/data/}
and consists of 44,484 training points and 4,449
testing points, which we then preprocess as follows.

We only use the first 3 features of the data and
remove the 73 duplicate points this projection causes.
Since the provided testing points overlap significantly with the training
points, we ignore the original testing points and instead randomly partition
the 44,411 remaining training points into 90\% training points (\( N = 39,969
\)) and 10\% prediction points (\( m = 4,442 \)).
Since the robot arm moves smoothly, the geometry of the dataset consists
of relatively continuous overlapping paths as shown in \cref{fig:sarcos}.
We use a Mat{\'e}rn kernel with smoothness \( \nu = 3/2 \) and length scale
\( \ell = 1 \), drawing \( 10^3 \) realizations from the resulting Gaussian
process for the target variable, ignoring the original objective to ensure
that the target is exactly Gaussian.

We consider three accuracy metrics for Gaussian process regression:
the log determinant of the posterior prediction covariance matrix,
the empirical coverage of the 90\% posterior prediction intervals
averaged over all realizations, and the average root mean square
error (RMSE) of the posterior means.
The log determinant is equivalent to the KL divergence by the
discussion in \cref{subsec:aggregated}, so the results are
similar to Cholesky factorization (\cref{subsec:chol_exp}).
We find that coverage is extremely accurate for all
methods (within \( 0.1\% \) for \( \rho > 2 \)).
Finally, the RMSE is shown in \cref{fig:gp_sarcos_rho}.

Despite an increased cost to select points, the conditional methods have
better accuracy per unit computational cost than their unconditional
counterparts as a result of their superior accuracy at the same sparsity.
However, the simpler method of \( k \)-NN also
achieves comparable accuracy per unit time cost.
As noted in \cite{gramacy2015speeding}, the simple method of
\( k \)-NN remains hard to beat without specialized tricks.

\begin{figure}[t]
  \centering
  \includegraphics{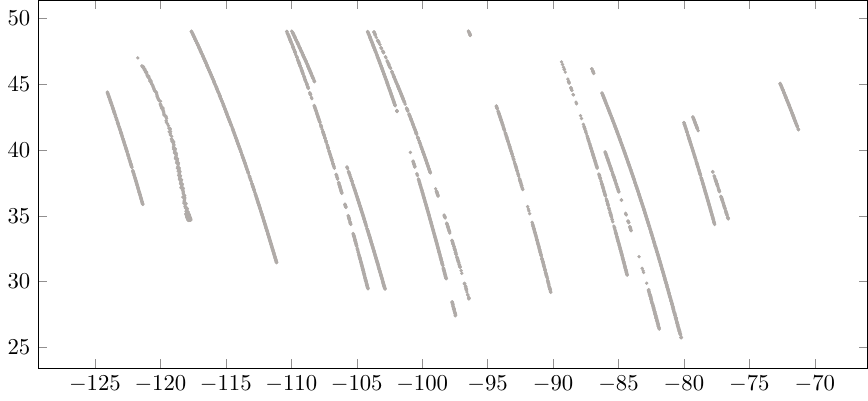}
  \caption{%
    Sampling 4,682 evenly spaced points from the OCO-2
    dataset, visualizing the first two features.
  }
  \label{fig:oco2}
\end{figure}

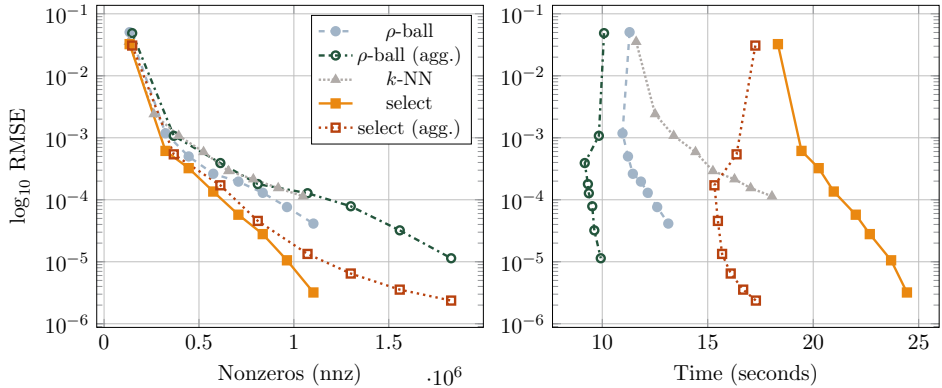
\begin{figure}[t]
  \centering
  \begin{tikzpicture}[baseline, scale=0.75]
  \begin{semilogyaxis}[
    grid={major},
    xlabel={Density (\( \mathsf{nnz} / N^2 \))},
    ylabel={RMSE},
    legend entries={%
      \( \rho \)-ball, \( \rho \)-ball (agg.),
      \( k \)-NN, select, select (agg.)
    },
    legend pos={north east},
    legend style={nodes={scale=0.9, transform shape}},
  ]
  \addplot table {figures/gp/oco2/data/rho_loss_KL.csv};
  \addplot table {figures/gp/oco2/data/rho_loss_KL_agg.csv};
  \addplot table {figures/gp/oco2/data/rho_loss_select-KNN.csv};
  \addplot table {figures/gp/oco2/data/rho_loss_select.csv};
  \addplot table {figures/gp/oco2/data/rho_loss_select_agg.csv};
  \end{semilogyaxis}
\end{tikzpicture}%
  \begin{tikzpicture}[baseline, scale=0.75]
  \begin{semilogyaxis}[
    grid={major},
    xlabel={Time (seconds)},
  ]
  \addplot table {figures/gp/oco2/data/rho_time_KL_loss.csv};
  \addplot table {figures/gp/oco2/data/rho_time_KL_agg_loss.csv};
  \addplot table {figures/gp/oco2/data/rho_time_select-KNN_loss.csv};
  \addplot table {figures/gp/oco2/data/rho_time_select_loss.csv};
  \addplot table {figures/gp/oco2/data/rho_time_select_agg_loss.csv};
  \end{semilogyaxis}
\end{tikzpicture}
  \caption{%
    We perform Gaussian process regression on the (modified)
    OCO-2 dataset (as visualized in \cref{fig:oco2}).
    The left panel shows the difference in RMSE from exact
    Gaussian process regression with varying density \( \rho \).
    The right panel shows the accuracy to computational
    time trade-off over varying \( \rho \).
  }
  \label{fig:gp_oco2_rho}
\end{figure}

\subsubsection{OCO-2 dataset}

We also use data produced by the OCO-2 project \cite{eldering2017orbiting}
at the Jet Propulsion Laboratory, California Institute of Technology,
obtained from the OCO-2 data archive maintained at the NASA Goddard Earth
Science Data and Information Services Center, which is available online at
\url{https://disc.gsfc.nasa.gov/datasets?keywords=oco2}.
OCO-2 is a polar-orbiting satellite designed
to measure atmospheric carbon dioxide.
The path an orbiting satellite takes creates characteristic
streaks in the data as shown in \cref{fig:oco2}.

We take data localized to the United States from the time
period 2017-05-16 to 2017-05-31, keeping only the features of
longitude, latitude, and time and removing any duplicate points.
We then take the first \( 2^{16} \) points and draw \( 10^3 \)
realizations from the Gaussian process with the same train test
split, kernel function, and hyperparameters as the SARCOS experiment.

As shown in \cref{fig:gp_oco2_rho}, the RMSE of \( k
\)-NN is worse than the ``\( \rho \)-ball'' baseline.
However, in nearly every other experiment (\cref{fig:chol_rho}
and \cref{fig:cg_iter}), \( k \)-NN does better than the
baseline, so neither method is strictly better than the other.
Our method, on the other hand, achieves the best accuracy (for a fixed
sparsity level) in \emph{every} numerical experiment we tried over a
wide variety of geometries and tasks, a testament to its robustness.

\subsection{Preconditioning for conjugate gradient}
\label{subsec:cg_exp}

Motivated by the equivalence of functionals like the Kaporin condition number
to the KL divergence, we investigate solving symmetric positive-definite
systems \( \CM \vec{x} = \vec{y} \) using the conjugate gradient and a sparse
Cholesky factor \( L \) as a preconditioner.
We note that from \cref{eq:kl} the KL divergence strongly
penalizes zero eigenvalues of the preconditioned matrix
\( \CM L L^{\top} \), improving its condition number.
In order to generate the covariance matrix \( \CM \) we sample up
to \( N = 2^{16} \) points uniformly at random from the unit cube
\( [0, 1]^3 \) and use a Mat{\'e}rn kernel with smoothness \( \nu
= 1/2 \) and length scale \( \ell = 1 \).
In exploratory numerical experiments, we found using higher smoothnesses
like \( \nu = 5/2 \) led to extremely poor condition numbers and
numerical instability (over thousands of iterations to converge).
Increasing length scale also worsens the
condition but to a less extreme extent.
Rather than generate a right-hand side \( \vec{y} \) directly, we first
sample a solution \( \vec{x} \sim \N(\vec{0}, \Id_N) \) and then compute
\( \vec{y} = \CM \vec{x} \) so that \( \vec{y} \) is realistically smooth.
When computing the preconditioner \( L \) by sparse Cholesky factorization,
we use a \( 2 \)-maximin ordering, a candidate set size scaling factor of
\( \rho_s = 2 \), and an aggregation parameter of \( \lambda = 1.5 \).
We then run the conjugate gradient algorithm with \( L \) as a preconditioner
until reaching a relative tolerance of \( 10^{-12} \).

To accelerate the matrix-vector product \( \CM \vec{x} \), both the
fast multipole method (e.g.\ \texttt{BBFMM3D} and \texttt{PBBFMM3D}
\cite{fong2009blackbox, wang2021pbbfmm3d}) and \( \mathcal{H} \)-matrix
approximations (e.g.\ \texttt{HLIBpro} \cite{borm2003introduction,
kriemann2005parallel, grasedyck2008parallel}) can be used.
In these experiments, we use \texttt{HLIBpro} with default settings
and a relative accuracy of \( \varepsilon = 10^{-6} \).

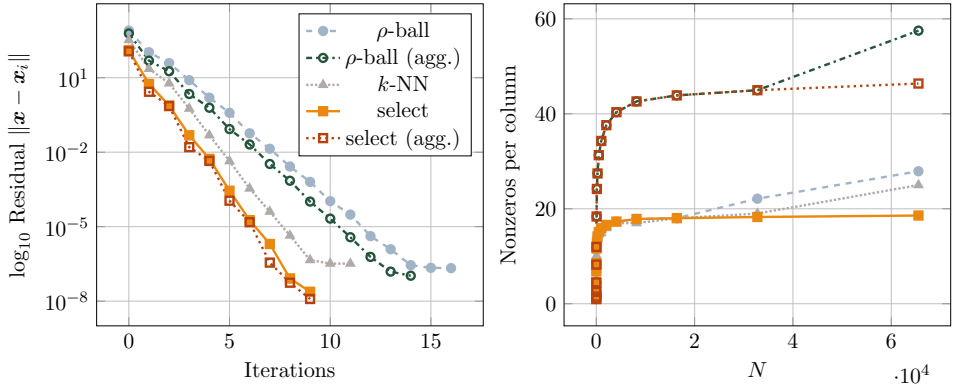
\begin{figure}[t]
  \centering
  \begin{tikzpicture}[baseline, scale=0.75]
  \begin{semilogyaxis}[
    grid={major},
    xlabel={Iterations},
    ylabel={Residual \( \norm{\vec{x} - \vec{x}_i} \)},
    legend entries={%
      \( \rho \)-ball, \( \rho \)-ball (agg.),
      \( k \)-NN, select, select (agg.), hlib
    },
    legend pos={north east},
    legend style={nodes={scale=0.9, transform shape}},
  ]
  \addplot table {figures/cg/data/n_iter-res_KL.csv};
  \addplot table {figures/cg/data/n_iter-res_KL_agg.csv};
  \addplot table {figures/cg/data/n_iter-res_select-KNN.csv};
  \addplot table {figures/cg/data/n_iter-res_select.csv};
  \addplot table {figures/cg/data/n_iter-res_select_agg.csv};
  \addplot table {figures/cg/data/n_iter-res_hlib.csv};
  \end{semilogyaxis}
\end{tikzpicture}%
  \begin{tikzpicture}[baseline, scale=0.75]
  \begin{axis}[
    grid={major},
    xlabel={\( N \)},
    ylabel={Nonzeros / bytesize per column},
  ]
  \addplot table {figures/cg/data/nnz_nnz_KL.csv};
  \addplot table {figures/cg/data/nnz_nnz_KL_agg.csv};
  \addplot table {figures/cg/data/nnz_nnz_select-KNN.csv};
  \addplot table {figures/cg/data/nnz_nnz_select.csv};
  \addplot table {figures/cg/data/nnz_nnz_select_agg.csv};
  \addplot table {figures/cg/data/nnz_nnz_hlib.csv};
  \end{axis}
\end{tikzpicture}
  \caption{%
    We use the conjugate gradient preconditioned by sparse Cholesky factors to
    solve the symmetric positive-definite system \( \CM \vec{x} = \vec{y} \).
    The left panel shows iteration progress for \( N =
    2^{16} \) points and a factor density of \( \rho = 4 \).
    The right panel shows the minimum number of nonzeros
    per column for conjugate gradient to converge within
    50 iterations with an increasing number of points.
  }
  \label{fig:cg_iter}
\end{figure}

When using \( \mathcal{H} \)-matrices as a preconditioner with
\texttt{HLIBpro}, we use the default settings with an admissibility
condition of \( \eta = 2 \) and set a relative accuracy of \( \varepsilon
= 10^{-\rho - 1.5} \) that is used for both the approximation for \(
\CM \) as well as its Cholesky factor.
The library \texttt{HLIBpro} only reports the size in bytes of the
resulting preconditioner; we compare this to number of nonzeros by
\( \mathsf{nnz} \approx \mathsf{bytesize} / 24 \) as each nonzero in
a Cholesky factor is represented by 24 bytes (a \texttt{float64} for
the value and two \texttt{int64}'s for the row and column indices).
A low relative accuracy can cause a loss of positive definiteness
and breakdown of the \( \mathcal{H} \)-Cholesky factorization.
If this happens, we add a nugget of \( \sigma^2 \Id \)
starting from \( \sigma^2 = 10^{-1} \) and multiplying by
\( 10^2 \) until the factorization runs to completion.
In practice, we do not exceed \( \sigma^2 = 10^1 \).
We report the total time taken by this procedure in \cref{fig:cg_iter}.

\begin{figure}[t]
  \centering
  \begin{tikzpicture}[baseline, scale=0.75]
  \begin{loglogaxis}[
    grid={major},
    xlabel={Density (\( \mathsf{nnz} / N^2 \))},
    ylabel={Iterations},
    legend entries={%
      \( \rho \)-ball, \( \rho \)-ball (agg.),
      \( k \)-NN, select, select (agg.), hlib
    },
    legend pos={north east},
    legend style={nodes={scale=0.7, transform shape}},
  ]
  \addplot table {figures/cg/data/rho_iter_KL.csv};
  \addplot table {figures/cg/data/rho_iter_KL_agg.csv};
  \addplot table {figures/cg/data/rho_iter_select-KNN.csv};
  \addplot table {figures/cg/data/rho_iter_select.csv};
  \addplot table {figures/cg/data/rho_iter_select_agg.csv};
  \addplot table {figures/cg/data/rho_iter_hlib.csv};
  \end{loglogaxis}
\end{tikzpicture}%
  \begin{tikzpicture}[baseline, scale=0.75]
  \begin{loglogaxis}[
    grid={major},
    xlabel={Density (\( \mathsf{nnz} / N^2 \))},
    ylabel={Wall-clock time (seconds)},
  ]
  \addplot table {figures/cg/data/rho_time_tot_KL.csv};
  \addplot table {figures/cg/data/rho_time_tot_KL_agg.csv};
  \addplot table {figures/cg/data/rho_time_tot_select-KNN.csv};
  \addplot table {figures/cg/data/rho_time_tot_select.csv};
  \addplot table {figures/cg/data/rho_time_tot_select_agg.csv};
  \addplot table {figures/cg/data/rho_time_tot_hlib.csv};
  \end{loglogaxis}
\end{tikzpicture}
  \caption{%
    The left panel shows how the number of iterations for the
    conjugate gradient to converge decreases with increasing
    preconditioner density \( \rho \) for \( N = 2^{16} \) points.
    The right panel shows the total wall-clock time (both to compute
    the preconditioner and to converge) with varying density.
  }
  \label{fig:cg_rho}
\end{figure}
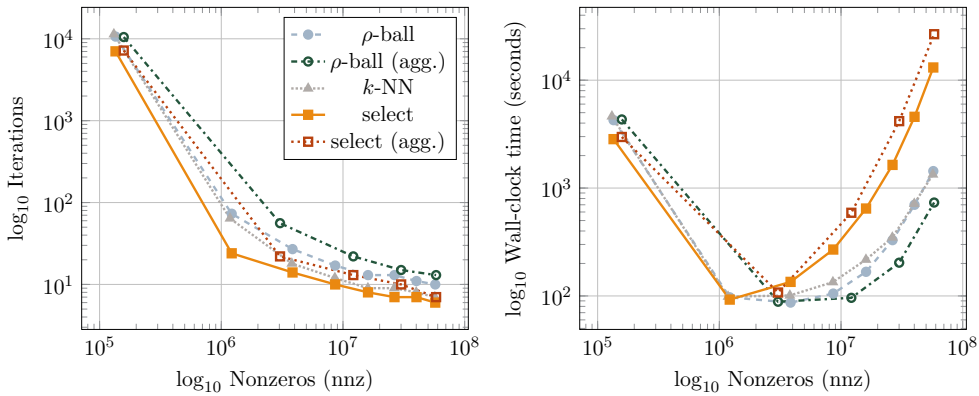

As shown in \cref{fig:cg_iter}, the conditional methods converge
in half the iterations of their unconditional counterparts
and aggregation barely reduces the number of iterations.
The minimum number of nonzeros per column to converge within a constant number
of iterations (here, 50) seems to grow logarithmically with the number of
points for all methods; although the conditional methods appear near constant.

We observe a characteristic ``U'' shape for the total wall-clock time
in \cref{fig:cg_rho} from the trade-off between spending computational
time in forming the preconditioner or in conjugate gradient iterations.
Across all preconditioner densities, the aggregated variants are slower
than their non-aggregated counterparts due to barely reducing the number
of iterations while producing significantly denser factors with slower
matrix-vector products.
The time-optimal density for conditional methods is sparser than
their unconditional counterparts due to using fewer iterations at
the same sparsity and it being more expensive to select nonzeros.
It is difficult to compare the total computational time of the methods, because
increasing the number of iterations (e.g., by demanding better tolerance or
using matrices with worse condition numbers) will prefer more accurate methods
even if they are slower to form the preconditioner.

For completeness, we report the operator norm of
\( \Id - \CM L L^\top \) in \cref{fig:cg_norm}.
We do not observe it to predict convergence well.
This is not surprising, since it is insensitive to \( L
L^\top \) failing to correct small eigenvalues of \( \CM \).
The latter are numerous, since \( \CM \)
is a discretization of a compact operator.

\pgfplotsset{
  cycle list={
    {very thick, lightblue, style=dashed},
    {very thick, orange,    style=solid},
  }
}

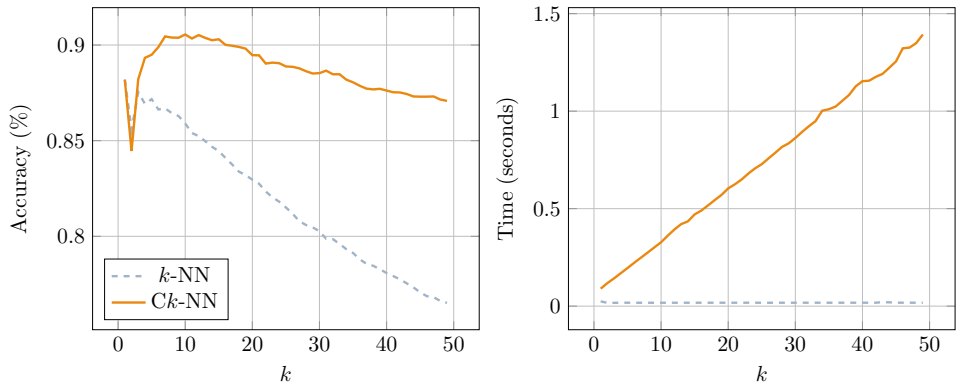
\begin{figure}[t]
  \centering
  \begin{tikzpicture}[baseline, scale=0.75]
  \begin{axis}[
    grid={major},
    xlabel={\( k \)},
    ylabel={Accuracy (\%)},
    legend entries={\( k \)-NN, C\( k \)-NN},
    legend pos={south west},
  ]
  \addplot table {figures/mnist/data/k_acc_scikit-knn.csv};
  \addplot table {figures/mnist/data/k_acc_conditional-knn.csv};
  \end{axis}
\end{tikzpicture}%
  \begin{tikzpicture}[baseline, scale=0.75]
  \begin{axis}[
    grid={major},
    xlabel={\( k \)},
    ylabel={Time (seconds)},
  ]
  \addplot table {figures/mnist/data/k_time_scikit-knn.csv};
  \addplot table {figures/mnist/data/k_time_conditional-knn.csv};
  \end{axis}
\end{tikzpicture}
  \caption{%
    We compare selection by Euclidean distance (\( k \)-NN) to
    conditional selection (C\( k \)-NN) in image classification.
    From the MNIST database \( N = 10^3 \) training images and \( m = 10^2
    \) testing images are randomly chosen; each test image is classified by
    the mode label of \( k \) selected images, and this process is repeated
    \( 10^2 \) times for each value of \( k \).
    (Left:) Accuracy with \( k \).
    (Right:) the time to select \( k \) points using C\( k \)-NN seems to
    scale linearly with \( k \) although it is quadratic asymptotically;
    possibly resulting from applying highly optimized BLAS operations to
    relatively small matrices.
  }
  \label{fig:mnist}
\end{figure}

\subsection{k-nearest neighbors selection}
\label{subsec:knn_exp}

We compare \( k \)-nearest neighbors (\( k \)-NN) to our selection
algorithm from \cref{subsec:greedy_select}, which we call conditional
\( k \)-nearest neighbors (C\( k \)-NN), in the following toy example
performing image classification.
From the MNIST database of handwritten digits
\cite{lecun1998gradientbased}, we randomly select \( N =
10^3 \) training images and \( m = 10^2 \) testing images.
We classify a test image by selecting \( k \) training
images and taking the mode label of those images.
For \( k \)-NN, we use the Euclidean distance and for C\(
k \)-NN, we use a Mat{\'e}rn kernel with smoothness \( \nu
= 3/2 \) and length scale \( \ell = 2^{10} \).
Accuracy is the percentage of test images classified correctly.

The accuracy of both methods decreases linearly with
increasing \( k \) as shown in \cref{fig:mnist}.
We hypothesize that nearby images are more likely to
have the same label, so selecting more points increases
the influence of further, differently labeled images.
C\( k \)-NN remains more accurate than \( k \)-NN for every \( k >
2 \), suggesting it consistently selects more informative images.
We emphasize that conditioning alone causes the difference in accuracy:
unconditional covariance decays monotonically with distance.

\subsection{Recovery of sparse Cholesky factors}
\label{subsec:recover_exp}

Motivated by the similarity of the selection algorithm to orthogonal matching
pursuit \cite{tropp2007signal}, we attempt to recover an \textit{a priori}
sparse Cholesky factor \( L \) from its precision matrix \( Q = L L^{\top} \).
To generate the nonzero entries of \( L \), for each column
we pick \( s \) lower triangular indices uniformly at
random and sample their values i.i.d.\ from \( \N(0, 1) \).
We fill \( L \)'s diagonal with a ``large'' positive
value 10 to ensure \( Q \) is well-conditioned.
The selection algorithm is given \( s \) and either \(
Q \) or the covariance matrix \( Q^{-1} \) depending on
which results in higher accuracy in reconstructing \( L \).
For a recovered sparsity pattern \( X \) and ground truth \( Y \) we report
\( \card{X \cap Y}/\card{X \cup Y} \), the intersection over union (IOU).
Using the KL divergence \cref{eq:L_obj} of the
recovered Cholesky factor is mostly equivalent to IOU.

As shown in \cref{fig:recover_acc}, C\( k \)-NN maintains a near-perfect
recovery accuracy much higher than the unconditional baselines.
In the setting of noisy measurements, noise sampled i.i.d.\ from
\( \N(0, \var) \) is symmetrically added to each entry of \( Q
\) (\( Q_{i, j} \) receives the same noise as \( Q_{j, i} \)).
Accuracy degrades with increasing noise for all methods, but C\( k \)-NN
is the most sensitive to noise as is shown in \cref{fig:recover_noise}.
At high enough levels of noise \( Q \) can lose
positive-definiteness, causing C\( k \)-NN to break down entirely.


\pgfplotsset{
  cycle list={
    {very thick, silver,    style=densely dotted},
    {very thick, lightblue, style=dashed},
    {very thick, seagreen,  style=dashdotted},
    {very thick, orange,    style=solid},
  }
}

\begin{figure}[t]
  \centering
  \begin{tikzpicture}[baseline, scale=0.75]
  \begin{axis}[
    grid={major},
    xlabel={\( N \)},
    ylabel={Accuracy (IOU)},
    ymin=0, ymax=1.1,
  ]
  \addplot table {figures/recover/data/n_score_random.csv};
  \addplot table {figures/recover/data/n_score_knn.csv};
  \addplot table {figures/recover/data/n_score_corr.csv};
  \addplot table {figures/recover/data/n_score_cknn.csv};
  \end{axis}
\end{tikzpicture}%
  \begin{tikzpicture}[baseline, scale=0.75]
  \begin{axis}[
    grid={major},
    xlabel={\( s \)},
    legend entries={rand., \( k \)-NN, corr., C\( k \)-NN},
    legend pos={south east},
    ymin=0, ymax=1.1,
  ]
  \addplot table {figures/recover/data/s_score_random.csv};
  \addplot table {figures/recover/data/s_score_knn.csv};
  \addplot table {figures/recover/data/s_score_corr.csv};
  \addplot table {figures/recover/data/s_score_cknn.csv};
  \end{axis}
\end{tikzpicture}
  \caption{%
    We attempt to recover a sparse Cholesky factor \(
    L \) of the precision matrix \( Q = L L^{\top} \).
    ``C\( k \)-NN'' minimizes the conditional variance of the target
    (diagonal) entry, ``corr.'' maximizes correlation with the target,
    ``\( k \)-NN'' maximizes covariance with the target, and ``rand.''
    samples entries uniformly at random.
    All methods achieve their best accuracy given \( Q \) except
    C\( k \)-NN which is given the covariance matrix \( Q^{-1} \).
    (Left:) Varying the size of \( L \), fixed density \( s = 2^5 \).
    (Right:) Varying \( s \), fixed size \( N = 2^8 \).
    Accuracy starts to improve from the factor nearing fully dense.
  }
  \label{fig:recover_acc}
\end{figure}
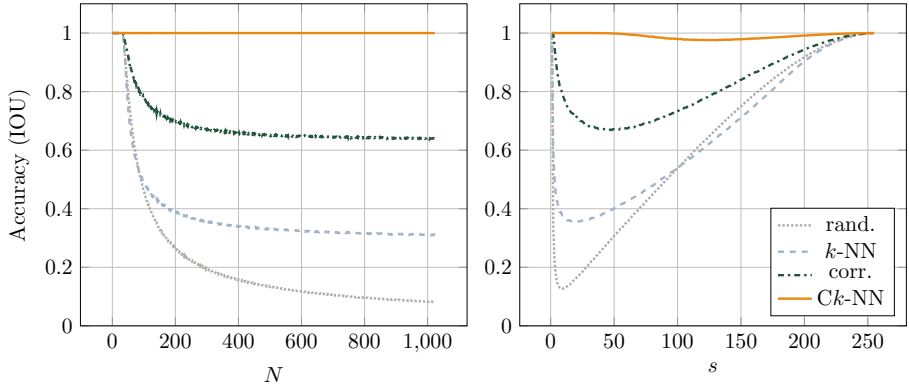

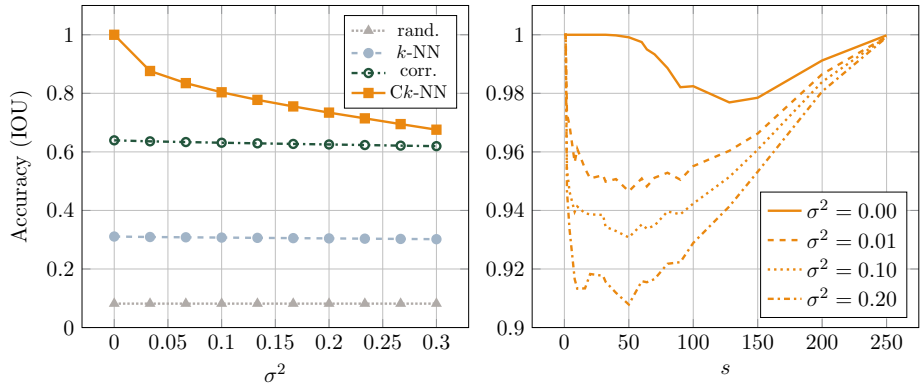
\begin{figure}[t]
  \centering
  \begin{tikzpicture}[baseline, scale=0.75]
  \begin{axis}[
    grid={major},
    every tick label/.append style={/pgf/number format/fixed},
    xlabel={\( \var \)},
    ylabel={Accuracy (IOU)},
    legend entries={rand., \( k \)-NN, corr., C\( k \)-NN},
    legend pos={north east},
    legend style={nodes={scale=0.85, transform shape}},
    ymin=0, ymax=1.1,
  ]
  \addplot+ [mark=triangle*]
    table {figures/recover/data/noise_score_random.csv};
  \addplot+ [mark=*]         table {figures/recover/data/noise_score_knn.csv};
  \addplot+ [mark=o]         table {figures/recover/data/noise_score_corr.csv};
  \addplot+ [mark=square*]   table {figures/recover/data/noise_score_cknn.csv};
  \end{axis}
\end{tikzpicture}%
  \begin{tikzpicture}[baseline, scale=0.75]
  \begin{axis}[
    grid={major},
    xlabel={\( s \)},
    legend entries={\( \var = 0.00 \), \( \var = 0.01 \),
      \( \var = 0.10 \), \( \var = 0.20 \)},
    legend pos={south east},
    ymin=0.9, ymax=1.01,
  ]
  \addplot [very thick, orange, style=solid]
    table {figures/recover/data/noise-s_score_cknn-0.00.csv};
  \addplot [very thick, orange, style=dashed]
    table {figures/recover/data/noise-s_score_cknn-0.01.csv};
  \addplot [very thick, orange, style=dotted]
    table {figures/recover/data/noise-s_score_cknn-0.10.csv};
  \addplot [very thick, orange, style=dashdotted]
    table {figures/recover/data/noise-s_score_cknn-0.20.csv};
  \end{axis}
\end{tikzpicture}
  \caption{%
    We add noise sampled i.i.d.\ from \( \N(0,
    \var) \) to the measurements \( Q \).
    The left panel shows accuracy with varying noise for \( N
    = 2^{10} \) columns and \( s = 2^5 \) nonzeros per column.
    The right panel shows accuracy for the C\( k \)-NN method at
    various noise levels for \( N = 2^8 \) and varying \( s \).
  }
  \label{fig:recover_noise}
\end{figure}

\section{Comparisons and conclusion}
\label{sec:conclusion}

We briefly compare our method to prior works.

\subsection{Comparison to other methods}
\label{subsec:compare}

\paragraph{Local approximate Gaussian process (laGP)}

Our conditional variance objective for point selection \cref{eq:obj_gp} was
first described in \cite{cohn1996neural} for optimal experimental design,
and has subsequently been named the active learning Cohn (ALC) technique.
\cite{gramacy2014local} and the follow-up works \cite{gramacy2015speeding,
sun2019emulating} apply ALC to directed Gaussian process inference, yielding
an algorithm equivalent to ours described in \cref{subsec:greedy_select} for
a single point of interest.
We note that their definition of conditional variance in Equation (5) is
analogous to our \cref{eq:cond_cov}, equating their conditional variance
objective in Equation (8) to our \cref{eq:obj_gp} (the connection is
explicitly stated in SM\S4's Equation (SM.22)).
They also update the precision by the blocked
equations in SM\S1 like our \cref{app:prec_insert}.

For inference at multiple points of interest, \cite{gramacy2014local} suggests
direct parallelization of the single-target algorithm over each target.
They mention that the \texttt{laGP} function in their
\texttt{R} package jointly considers multiple target
points, but without providing details on this procedure.
\cite{sun2019emulating} proposes a ``joint'' or ``path'' ALC by taking
the average reduction in posterior variance over target points.
Instead, we generalize ALC to multiple prediction points through
their posterior log determinant \cref{eq:greedy_mult} and provide
an explicit algorithm described in \cref{subsec:mult_select}.
In addition, integration of the algorithm into Cholesky factorization
(\cref{sec:chol_select}) provides a global approximation of the Gaussian
process (\cref{subsec:gp_exp}) beyond directed local approximation.

\paragraph{%
  Fast multipole method, \( \mathcal{H}
  \)-matrices, and Nystr{\"o}m approximation%
}
The Nystr{\"o}m approximation \cite{williams2000using,
chen2022randomly} is a low-rank approximation often formed
from the first \( k \) columns of the Cholesky factor of \( \CM \).
This amounts to assuming that conditioned on the first
\( k \) variables, the process is deterministic.
Thus, our method generalizes the Nystr{\"o}m approximation
to more general conditional independence structures.
Global low-rank approximations like Nystr{\"o}m
oversmooth in multi-scale problems arising, for instance,
in spatial statistics \cite{stein2014limitations}.
To overcome this limitation, tree codes
\cite{barnes1986hierarchical,march2016askit}, panel clustering
\cite{hackbusch1986complexity,hackbusch1989fast} and fast multipole methods
\cite{rokhlin1985rapid,greengard1987fast,ying2004kernel,fong2009blackbox,
wang2021pbbfmm3d} apply low-rank approximations hierarchically
to long range interactions, often exploiting smoothness of the
kernel function to devise polynomial approximants.
This allows computing matrix-vector products \( \CM \vec{x}
\) at near-linear complexity but \( \CM^{-1} \vec{x} \)
requires iterative solvers that may converge slowly.
So-called inverse fast multipole methods aim to circumvent
this problem \cite{ambikasaran2014inverse,coulier2017inverse}.
Our method can be used in tandem with a FMM in a
preconditioned conjugate gradient to compute inverses,
potentially improving upon \cite{chen2024lightning}.
\( \mathcal{H} \)-matrices \cite{borm2003introduction,kriemann2005parallel,
grasedyck2008parallel,kriemann2013lu,hackbusch2015hierarchical}, are
an abstraction of the fast multipole method that decomposes matrices
into low-rank blocks. The \( \mathcal{H} \) format supports a Cholesky
factorization on kernel matrices.
See \cite{litvinenko2019hlibcov,geoga2020scalable} for
examples of its application to Gaussian process regression.
The \( \mathcal{H} \) format captures a strictly larger set
of covariance matrices than the sparse inverse--Cholesky
factorization of \cite{schafer2021sparse} and this work.
But \( \mathcal{H} \) matrix--based approaches are prone to losing
positive-definiteness (see \cref{subsec:cg_exp}), complicated, and their
near-linear asymptotic complexity is offset by large constant factors.
\cite{schafer2021sparse} shows that for general elliptic Green's matrices,
the rigorous asymptotic error-vs-cost trade-off of sparse inverse Cholesky
factorization improves over that of \( \mathcal{H} \) (or \( \mathcal{H}^2
\)) matrices.
But fast multipole methods and \( \mathcal{H} \)-matrices efficiently
implement accurate matrix-vector products in many practical settings. This
motivates combining them with sparse inverse Cholesky preconditioners as
discussed in \cite{chen2024lightning} and \cref{subsec:cg_exp}.

\paragraph{Orthogonal matching pursuit (OMP)}

Our selection algorithm can be viewed as the covariance equivalent
of the sparse signal recovery algorithm orthogonal matching
pursuit (OMP) \cite{tropp2007signal, tropp2006algorithms}.
OMP measures the approximation of a target signal by its residual after
orthogonal projection onto the subspace of chosen training signals,
which is efficiently calculated by maintaining a QR factorization.
In contrast, our selection algorithm uses a kernel function to evaluate inner
products, so orthogonalization becomes conditioning, the residual norm
becomes variance, and the QR factorization becomes a Cholesky factorization.
A difference from OMP is that the computational time of conditioning
dominates that of evaluating the kernel function since the feature
space is low-dimensional (often 2 or 3 in spatial statistics).

\paragraph{Sparse Cholesky factorization}

Our sparse Cholesky factorization algorithm proposed in
\cref{sec:chol_select} relies heavily on the KL-minimization framework of
\cite{schafer2021sparse} and is similar to \cite{kang2021correlationbased}.
We comment that using \( k \)-nearest neighbors to select the
sparsity set instead of our conditional selection algorithms
essentially recovers \cite{schafer2021sparse} and that using
the correlation objective \cref{eq:obj_gp} without conditioning
on selected points recovers \cite{kang2021correlationbased}.

\subsection{Conclusion}

In this work, we develop an algorithm for directed Gaussian process
regression, which greedily selects training points that maximize
mutual information with a target point, conditional on all points
previously selected to avoid redundancy.
We show that using conditional selection to pick the sparsity pattern of
sparse approximate Cholesky factors of precision matrices significantly
improves accuracy and performance in downstream tasks compared to selection
by nearest neighbors.
Single-target conditional selection is computationally efficient and can
be extended to the settings of multiple-target and partial conditioning
corresponding to aggregated (or supernodal) Cholesky factorization.
Finally, global selection gives a principled way of
distributing nonzeros over columns of the Cholesky factor.
We support these claims through extensive numerical
experimentation in a variety of problems.

\section*{Acknowledgments}

This research was supported in part by research cyberinfrastructure resources
and services provided by the Partnership for an Advanced Computing Environment
(PACE) at the Georgia Institute of Technology, Atlanta, Georgia, USA.
HO acknowledges support from the Air Force Office of Scientific Research
under MURI award number FA9550-20-1-0358 (Machine Learning and Physics-Based
Modeling and Simulation) and from the Department of Energy under award number
DE-SC0023163 (SEA-CROGS: Scalable, Efficient and Accelerated Causal Reasoning
Operators, Graphs and Spikes for Earth and Embedded Systems).
MK was partially supported by National
Science Foundation (NSF) Grant DMS--1953005.
SH and FS gratefully acknowledge support from the Office of Naval
Research under grant N00014-23-1-2545 (PM Dr.\ Reza Malek-Madani).
JG was partially supported by the National Science Foundation
under grant numbers DMS--1916208 and DMS--1953088.
FS thanks Joel Tropp for fruitful discussions on JT's proposal to use matching
pursuit algorithms for selecting sparsity patterns of Cholesky factors.

\bibliographystyle{siamplain}
\bibliography{references}

\newpage

\appendix

\section{Derivations in KL-minimization}

\subsection{KL divergence of optimal factor}
\label{app:kl_L}

\begin{proof}[Proof of Equation \cref{eq:obj_chol}]
  From the closed-form expression for the KL divergence in
  \cref{eq:kl} and defining \( \Delta \defeq 2 \KL*{\N(\vec{0},
  \CM)}{\N(\vec{0}, (L L^{\top})^{-1})} \) for brevity of notation,
  \begin{align}
    \label{eq:kl_L}
    \Delta &= \trace(L L^{\top} \CM) - \logdet(L L^{\top}) - \logdet(\CM) - N.
    \shortintertext{
      Focusing on the term \( \trace(L L^{\top} \CM) = \trace(L^{\top} \CM L)
      \) by the cyclic property of trace and using the sparsity of \( L \) by
      plugging in the definition \cref{eq:L_col} for each column \( L_{s_i, i}
      \),
    }
    \trace(L^{\top} \CM L) &= \sum_{i = 1}^N
      L_{s_i, i}^{\top} \CM_{s_i, s_i} L_{s_i, i} \\
    &= \sum_{i = 1}^N
      \left (
        \frac{\left ( \CM_{s_i, s_i}^{-1} \vec{e}_1 \right )^{\top}}
          {\sqrt{\vec{e}_1^{\top} \CM_{s_i, s_i}^{-1} \vec{e}_1}}
      \right )
      \CM_{s_i, s_i}
      \left (
        \frac{\CM_{s_i, s_i}^{-1} \vec{e}_1}
          {\sqrt{\vec{e}_1^{\top} \CM_{s_i, s_i}^{-1} \vec{e}_1}}
      \right ) \\
    &= \sum_{i = 1}^N
      \frac{\vec{e}_1^{\top} \CM_{s_i, s_i}^{-1}
            \CM_{s_i, s_i} \CM_{s_i, s_i}^{-1}
            \vec{e}_1}{\vec{e}_1^{\top} \CM_{s_i, s_i}^{-1} \vec{e}_1}
    = \sum_{i = 1}^N 1 = N,
    \shortintertext{
      exactly the constraint \( \diag(L^{\top} \CM L) = 1 \)
      from \cref{subsec:vecchia}.
      Substituting back into \cref{eq:kl_L},
    }
    \Delta &= -\logdet(L L^{\top}) - \logdet(\CM).
    \shortintertext{
      Computing the log determinant of a triangular matrix as
      the sum of the log of its diagonal entries and plugging in
      the definition \cref{eq:L_col} for the diagonal entries,
    }
    &= -\sum_{i = 1}^N
      \left [ \log(\vec{e}_1^{\top} \CM_{s_i, s_i}^{-1} \vec{e}_1) \right ]
      - \logdet(\CM) \\
    &= \sum_{i = 1}^N
      \left [
        \log
        \left (
          (\vec{e}_1^{\top} \CM_{s_i, s_i}^{-1} \vec{e}_1)^{-1}
        \right )
      \right ]
      - \logdet(\CM).
    \shortintertext{
      Now we use that conditioning in
      covariance is marginalization in precision,
    }
    \label{eq:inverse_cond}
    \CM_{1, 1 \mid 2} &= \left ( \CM^{-1} \right )_{1, 1}^{-1}
    \qquad \qquad \text{for }
    \CM =
      \begin{pmatrix}
        \CM_{1, 1} & \CM_{1, 2} \\
        \CM_{2, 1} & \CM_{2, 2}
      \end{pmatrix}.
    \shortintertext{
      Transforming the marginalization \( (\vec{e}_1^{\top} \CM_{s_i, s_i}^{-1}
      \vec{e}_1)^{-1} = \left (\CM_{s_i, s_i}^{-1} \right )_{1, 1}^{-1} =
      \CM_{i, i \mid s_i \setminus \{ i \}} \) by \cref{eq:inverse_cond},
    }
    \Delta &= \sum_{i = 1}^N
      \left [
        \log \left ( \CM_{i, i \mid s_i \setminus \{ i \}} \right )
      \right ]
      - \logdet(\CM). \\
    \shortintertext{Now we use the chain rule of log determinants,
      using the same blocking as \cref{eq:inverse_cond},}
    \label{eq:det_chain}
    \logdet(\CM) &= \logdet(\CM_{1, 1}) + \logdet(\CM_{2, 2 \mid 1}). \\
    \shortintertext{
      Repeatedly expanding the log determinant by
      \cref{eq:det_chain}, working from back to front,
    }
    \Delta &= \sum_{i = 1}^N
        \log \left ( \CM_{i, i \mid s_i \setminus \{ i \}} \right ) -
      \sum_{i = 1}^N
        \log \left ( \CM_{i, i \mid i + 1:} \right ) \\
    &= \sum_{i = 1}^N
      \left [
        \log \left ( \CM_{i, i \mid s_i \setminus \{ i \}} \right ) -
        \log \left ( \CM_{i, i \mid i + 1:} \right )
      \right ].
  \end{align}
\end{proof}

\subsection{Aggregated KL divergence}
\label{app:kl_agg}

\begin{proof}[Proof of Equation \cref{eq:obj_mult}]
  The KL divergence \cref{eq:obj_chol}
  restricted to the group \( \tilde{i} \) is
  \begin{align}
    \sum_{i \in \tilde{i}} \log(\CM_{i \mid s_i \setminus \{ i \} }) &=
      \log(\CM_{i_1 \mid \tilde{s}}) +
      \log(\CM_{i_2 \mid \tilde{s} \cup \{ i_1 \}}) + \dotsb +
      \log(\CM_{i_m \mid \tilde{s} \cup \tilde{i}}) \\
    \shortintertext{
      where we write \( \CM_j \defeq \CM_{j, j} \).
      Combining the first two terms by the chain rule \cref{eq:det_chain},
    }
    &= \logdet(\CM_{\{ i_1, i_2 \} \mid \tilde{s}}) +
      \log(\CM_{i_3 \mid \tilde{s} \cup \{ i_1, i_2 \}}) + \dotsb +
      \log(\CM_{i_m \mid \tilde{s} \cup \tilde{i}}). \\
    \shortintertext{
      Proceeding by induction, we are able to
      reduce the entire sum to the single term
    }
    &= \logdet(\CM_{\tilde{i}, \tilde{i} \mid \tilde{s}}).
  \end{align}
\end{proof}

\subsection{Partial KL divergence}
\label{app:partial}

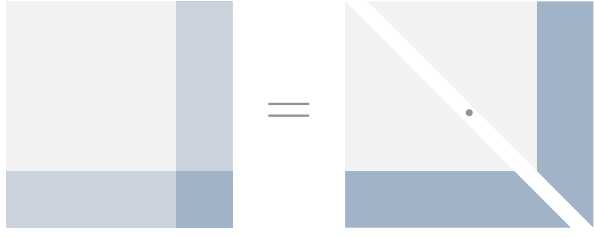
\begin{figure}[t]
  \centering
  \begin{tikzpicture}[scale=3]
  \fill[lightsilver, opacity=1.0]   (0,        0) rectangle (1, -1);
  \fill[lightblue,   opacity=0.5]   (0,    -0.75) rectangle (0.75, -1);
  \fill[lightblue,   opacity=0.5]   (0.75,     0) rectangle (1, -0.75);
  \fill[lightblue,   opacity=1.0]   (0.75, -0.75) rectangle (1, -1);

  \node at (1.25, -0.5) {\Huge \textcolor{darksilver}{$=$}};

  \fill[lightsilver]
    (1.5, -1) -- (2.5, -1) -- (1.5, 0) -- cycle;
  \fill[lightblue]
    (1.5, -1) -- (2.5, -1) -- (2.25, -0.75) -- (1.5, -0.75) -- cycle;

  \node at (2.05, -0.5) {\Huge \textcolor{darksilver}{$\cdot$}};

  \fill[lightsilver]
    (2.6, -1) -- (2.6, 0) -- (1.6, 0) -- cycle;
  \fill[lightblue]
    (2.6, -1) -- (2.6, 0) -- (2.35, 0) -- (2.35, -0.75) -- cycle;
\end{tikzpicture}
  \caption{%
    Illustration of the Cholesky factorization of
    a partially conditioned covariance matrix.
    Here \textcolor{darksilver}{grey} denotes fully unconditional,
    \textcolor{darklightblue}{blue} denotes fully conditional, and the
    \textcolor{silver!50!lightblue}{mixed color} denotes interaction
    between the two.
    Surprisingly, such a matrix factors into a ``pure'' Cholesky
    factor by ``gluing'' the prefix of the fully unconditional
    factor with the suffix of the fully conditional factor.
  }
  \label{fig:partial_factor}
\end{figure}

\begin{proof}[Proof of Equation \cref{eq:partial_kl}]
  The original variables \( \vec{y} \) have joint density multivariate
  Gaussian, \( \vec{y} \sim \N(\vec{0}, \CM) \) for some covariance matrix
  \( \CM \) so the fully conditional variables \( \vec{y}_{\mid k} \) have
  posterior distribution from \cref{eq:cond_mean} and \cref{eq:cond_cov},
  \( \vec{y}_{\mid k} \sim \N(\vec{\mean}, \CM_{:, : \mid k}) \) for some
  posterior mean \( \vec{\mean} \).
  The covariance of unconditioned \( y_i \) and \( y_j \) is \( \CM_{i,
  j} \) by definition; similarly, the covariance of conditioned \(
  y_{i \mid k} \) and \( y_{j \mid k} \) is \( \CM_{i, j \mid k} \).
  We must compute the covariance between unconditioned
  \( y_i \) and conditioned \( y_{j \mid k} \).
  Let \( L = \chol(\CM) \) and \( L' = \chol(\CM_{:, : \mid k}) \)
  so that \( \vec{y} = L \vec{z} \) for \( \vec{z} \sim \N(\vec{0},
  \Id) \) and \( \vec{y}_{\mid k} = L' \vec{z} + \vec{\mean} \).
  By the definition of covariance,
  \begin{align}
    \Cov[y_i, y_{j \mid k}] &=
      \E[(y_i - \E[y_i])(y_{j \mid k} - \E[y_{j \mid k}])]
      = \E[(L_i \vec{z}) ({L'}_j \vec{z} + \mu_j - \mu_j)] \\
    &= \E[(L_{i, 1} z_1 + \dotsb + L_{i, m} z_m)
          ({L'}_{j, 1} z_1 + \dotsb + {L'}_{j, m} z_m)].
    \shortintertext{
      For \( i \neq j \), \( \E[z_i z_j] = \E[z_i] \E[z_j] = 0 \)
      since \( z_i \) is independent of \( z_j \) and has mean 0,
    }
    &= L_{i, 1} {L'}_{j, 1} \E[z_1^2] + \dotsb +
       L_{i, m} {L'}_{j, m} \E[z_m^2].
    \shortintertext{
      For any \( i \), \( \E[z_i^2] = \Var[z_i] + \E[z_i]^2 = 1 + 0 = 1 \),
    }
    &= L_{i, 1} {L'}_{j, 1} + \dotsb + L_{i, m} {L'}_{j, m}
    = L_i^{\top} {L'}_j.
    \shortintertext{
      Thus, the new covariance matrix factors into
      two Cholesky factors ``glued'' together,
    }
    \Cov[\vec{y}_{\parallel k}] &=
    \begin{pmatrix}
      L_{:p} L_{:p}^{\top} &
      L_{:p} {L'}_{p + 1:}^{\top} \\
      {L'}_{p + 1:} L_{:p}^{\top} &
      {L'}_{p + 1:} {L'}_{p + 1:}^{\top}
    \end{pmatrix} =
    \begin{pmatrix}
      L_{:p} \\
      {L'}_{p + 1:}
    \end{pmatrix}
    \begin{pmatrix}
      L_{:p} \\
      {L'}_{p + 1:}
    \end{pmatrix}^{\top}
  \end{align}
  which is illustrated in \cref{fig:partial_factor}.
  Armed with this representation, we equate the log determinant of \(
  \CM_{\tilde{i}, \tilde{i} \parallel k} \defeq \Cov[\vec{y}_{\parallel
  k}] \) to the KL divergence in \cref{eq:obj_chol}.
  Recalling that the determinant of a triangular
  matrix is the product of its diagonal entries,
  \begin{align}
    \nonumber
    \frac{1}{2} \logdet(\CM_{\tilde{i}, \tilde{i} \parallel k}) &=
    \underbrace{
      \log(L_{1, 1}) + \dotsb + \log(L_{p, p})
    }_{\text{the same}} +
    \underbrace{
      \log({L'}_{p + 1, p + 1}) + \dotsb + \log({L'}_{m, m})
    }_{\text{conditioned}}.
  \end{align}
  Comparing to the KL divergence \cref{eq:obj_chol} and
  recalling that \( k \) is added to \( s_i \) if \( i > p \),
  \begin{align}
    \sum_{i = 1}^m \log \left ( \CM_{i, i \mid s_i \setminus \{ i \}} \right )
    = &\underbrace{
        \log \left ( \CM_{1, 1 \mid s_1 \setminus \{ 1 \}} \right ) + \dotsb +
        \log \left ( \CM_{p, p \mid s_p \setminus \{ p \}} \right )
      }_\text{the same} + \\
    \nonumber
    &\underbrace{
      \log \left (
        \CM_{p + 1, p + 1 \mid s_{p + 1} \setminus \{ p + 1 \}}
      \right ) + \dotsb +
      \log \left ( \CM_{m, m \mid s_m \setminus \{ m \}} \right )
    }_\text{conditioned}.
    \shortintertext{
      Since \( L_{i, i} \) (and \( {L'}_{i, i} \)) is the square root
      of the posterior variance of the \( i \)-th variable from the
      statistical perspective in \cref{eq:chol}, we have \( 2 \log(L_{i,
      i}) = \log(\CM_{i, i \mid s_i \setminus \{ i \}}) \) and so
    }
    \logdet(\CM_{\tilde{i}, \tilde{i} \parallel k}) &=
      \sum_{i = 1}^m
        \log \left ( \CM_{i, i \mid s_i \setminus \{ i \}} \right ).
  \end{align}
\end{proof}

\subsection{Aggregated computation of sparsity entries}
\label{app:L_mult}

We wish to compute the entries of the sparse Cholesky factor
\( L_{s_i, i} \) according to \cref{eq:L_col} in the aggregated
sparsity setting of \cref{subsec:aggregated}.
Recall that we have aggregated the column indices \( i_1 \succ
\dotsb \succ i_m \) into \( \tilde{i} = \{ i_1, \dotsc, i_m \} \)
and \( s_{\tilde{i}} \) denotes the aggregated sparsity pattern.
The sparsity pattern for the \( i \)-th column of the group is all
those entries that satisfy lower triangularity, \( s_i \defeq \{
j \in s_{\tilde{i}} : j \succeq i \} \subseteq s_{\tilde{i}} \).
Because each column's sparsity pattern is a subset of the
overall sparsity pattern, it is possible to compute an outer
approximation and specialize to each column efficiently.
Assuming the sparsity entries \( s_{\tilde{i}} \) are sorted according to
\( \prec \), let \( Q \defeq \CM_{s_{\tilde{i}}, s_{\tilde{i}}}^{-1} \) be
the precision of the aggregated sparsity pattern and let \( k \) be the \(
i \)-th column's index in \( s_{\tilde{i}} \).
Because \( s_{\tilde{i}} \) is sorted according to \(
\prec \), the sparsity pattern for the \( i \)-th column
\( s_i \) is exactly the entries \( k \) and after,
\begin{align}
  \label{eq:L_precision}
  L_{s_i, i}
   &= \frac{\CM_{s_i, s_i}^{-1} \vec{e}_1}
           {\sqrt{\vec{e}_1^{\top} \CM_{s_i, s_i}^{-1} \vec{e}_1}}
    = \frac{\left (Q^{-1} \right )_{k:, k:}^{-1} \vec{e}_1}
           {\sqrt{\vec{e}_1^{\top} \left (Q^{-1} \right )_{k:, k:} \vec{e}_1}}
    = \frac{Q_{k:, k: \mid :k - 1} \vec{e}_1}
           {\sqrt{\vec{e}_1^{\top} Q_{k:, k: \mid :k - 1} \vec{e}_1}}
\end{align}
where the first equality follows from \( Q^{-1} =
\CM_{s_{\tilde{i}}, s_{\tilde{i}}} \) and \( s_i \) is
the \( k \)-th index of \( s_{\tilde{i}} \) and after.
We turn marginalization in precision into
conditioning in covariance by \cref{eq:inverse_cond}.

So we want the \( k \)-th column of \( Q
\), conditional on all columns before it.
From \cref{eq:chol}, this can be directly read off the \(
k \)-th column of the Cholesky factor \( L = \chol(Q) \)
to compute \cref{eq:L_col} for each \( i \in \tilde{i} \).
However, computing \( Q = \CM_{s_{\tilde{i}}, s_{\tilde{i}}}^{-1} \) by
inverting \( \CM_{s_{\tilde{i}}, s_{\tilde{i}}} \) and then additionally
computing its Cholesky factor \( L = \chol(Q) \) is a bit wasteful.

Instead of computing a \emph{lower} triangular factor for the precision,
we can compute an \emph{upper} triangular factor for the covariance whose
inverse transpose will be a lower triangular factor for the precision.
Let \( U = P^{\Reverse} \chol(P^{\Reverse} \CM_{s_{\tilde{i}},
s_{\tilde{i}}} P^{\Reverse}) P^{\Reverse} \) where \( P^{\Reverse} \) is the
order-reversing permutation; \( U \) is upper triangular and satisfies \( U
U^{\top} = \CM_{s_{\tilde{i}}, s_{\tilde{i}}} \) so \( U^{-\top} U^{-1} =
\CM_{s_{\tilde{i}}, s_{\tilde{i}}}^{-1} = Q \) and we see that \( L = U^{-\top}
\) is a lower triangular factor satisfying \( L L^{\top} = Q \).
Thus we can compute \( U \) as a Cholesky factor of the covariance
and dynamically form the \( k \)-th column of \( L \) by solving
the triangular system \( L_{:, k} = U^{-\top} \vec{e}_k \).
This method is described in Algorithm 3.2, Figure
4, and Appendix A.2 of \cite{schafer2021sparse}.


\section{Computation in sparse Gaussian process selection}

\subsection{Mutual information objective}
\label{app:mutual_info}

The \emph{mutual information} or \emph{information
gain} between two collections of random variables \(
\vec{y}_\Pred \) and \( \vec{y}_\Train \) is defined as
\begin{align}
  \label{eq:info}
  \MI[\vec{y}_\Pred;\vec{y}_\Train] &\defeq \entropy[\vec{y}_\Pred] -
    \entropy[\vec{y}_\Pred \mid \vec{y}_\Train].
\end{align}
Maximizing the mutual information is equivalent to minimizing the
conditional entropy since the entropy of \( \vec{y}_\Pred \) is constant.
Because the differential entropy of a multivariate Gaussian monotonically
increases with the determinant of its covariance matrix, minimizing the
conditional entropy is equivalent to minimizing the determinant of the
posterior covariance matrix.
The determinant of a single prediction point is its variance so we
can minimize the \emph{conditional variance} of the target point.

Supposing \( \vec{y}_\Pred \) is estimated by the posterior
mean \cref{eq:cond_mean}, the estimator is unbiased because \(
\E[\E[\vec{y}_\Pred \mid \vec{y}_\Train]] = \E[\vec{y}_\Pred] \) and
so the expected mean squared error of the estimator \( \text{MSE}
\defeq \E[(\vec{y}_\Pred - \E[\vec{y}_\Pred \mid \vec{y}_\Train])^2]
\) is simply the conditional variance because
\begin{align}
  \text{MSE}
  = \E[\E[(\vec{y}_\Pred - \E[\vec{y}_\Pred \mid \vec{y}_\Train])^2
          \mid \vec{y}_\Train]]
  = \E[\Var[\vec{y}_\Pred \mid \vec{y}_\Train]]
  = \Var[\vec{y}_\Pred \mid \vec{y}_\Train],
\end{align}
from the quirk that the posterior covariance \cref{eq:cond_cov}
does not depend on observing  \( \vec{y}_\Train \).

So maximizing the mutual information is equivalent to minimizing the
posterior variance (or log determinant) which is in turn equivalent
to minimizing the expected mean squared error of the prediction.
Yet another perspective arises from comparing the definition of mutual
information \cref{eq:info} to the EV-VE identity \cref{eq:eve},
\begin{align}
  \label{eq:info_eve}
  \textcolor{darkorange}{\entropy[\vec{y}_\Pred]} &=
    \textcolor{lightblue}{\entropy[\vec{y}_\Pred \mid \vec{y}_\Train]} +
         \textcolor{rust}{\MI[\vec{y}_\Pred;\vec{y}_\Train]} \\
  \label{eq:eve}
  \textcolor{darkorange}{\Var[\vec{y}_\Pred]} &=
    \textcolor{lightblue}{\E[\Var[\vec{y}_\Pred \mid \vec{y}_\Train]]} +
         \textcolor{rust}{\Var[\E[\vec{y}_\Pred \mid \vec{y}_\Train]]}
\end{align}
On the left hand side, \textcolor{darkorange}{variance}
is monotone with \textcolor{darkorange}{entropy}.
On the right-hand side, the \textcolor{lightblue}{conditional
variance} is monotone with \textcolor{lightblue}{conditional entropy}.
Because the two terms on the right-hand side add to a constant, minimizing
the \textcolor{lightblue}{conditional variance} is equivalent to maximizing
the \textcolor{rust}{variance of conditional expectation}, which, by process
of elimination, corresponds to the \textcolor{rust}{mutual information}.
We can intuitively interpret the variance of the conditional expectation
as measuring the extent to which the estimator for \( \vec{y}_\Pred \),
the conditional expectation \( \E[\vec{y}_\Pred \mid \vec{y}_\Train] \),
varies with observed values of \( \vec{y}_\Train \).

\subsection{Cholesky factorization as iterative conditioning}
\label{app:chol_stat}

Factoring the joint covariance matrix \( \CM \) blocked as
\cref{eq:inverse_cond} by two steps of block Gaussisan elimination,
\begin{align}
  \nonumber
  \begin{pmatrix}
    \CM_{1, 1} & \CM_{1, 2} \\
    \CM_{2, 1} & \CM_{2, 2}
  \end{pmatrix} &=
  \begin{pmatrix}
    \Id & 0 \\
    \textcolor{darkorange}{\CM_{2, 1} \CM_{1, 1}^{-1}} & \Id
  \end{pmatrix}
  \begin{pmatrix}
    \CM_{1, 1} & 0 \\
    0 & \textcolor{lightblue}{
      \CM_{2, 2} - \CM_{2, 1} \CM_{1, 1}^{-1} \CM_{1, 2}
    }
  \end{pmatrix}
  \begin{pmatrix}
    \Id & \textcolor{darkorange}{\CM_{1, 1}^{-1} \CM_{1, 2}} \\
    0 & \Id
  \end{pmatrix}
  \shortintertext{
    so we see that the Cholesky factorization
    of the joint covariance \( \CM \) is
  }
  \chol(\CM) &=
  \begin{pmatrix}
    \Id & 0 \\
    \textcolor{darkorange}{\CM_{2, 1} \CM_{1, 1}^{-1}} & \Id
  \end{pmatrix}
  \begin{pmatrix}
    \chol(\CM_{1, 1}) & 0 \\
    0 & \chol(\textcolor{lightblue}{
      \CM_{2, 2} - \CM_{2, 1} \CM_{1, 1}^{-1} \CM_{1, 2}
    })
  \end{pmatrix} \\
  \label{eq:chol}
  &=
  \begin{pmatrix}
    \textcolor{darkorange}{\chol(\CM_{1, 1})} & 0 \\
    \textcolor{darkorange}{\CM_{2, 1} \chol(\CM_{1, 1})^{-\top}} &
    \chol(\textcolor{lightblue}{
      \CM_{2, 2} - \CM_{2, 1} \CM_{1, 1}^{-1} \CM_{1, 2}
    })
  \end{pmatrix}
\end{align}
where the conditional expectation \cref{eq:cond_mean} corresponds
to \( \textcolor{darkorange}{\CM_{2, 1} \CM_{1, 1}^{-1}} \) and
the conditional covariance \cref{eq:cond_cov} corresponds to
\(
  \textcolor{lightblue}{\CM_{2, 2} - \CM_{2, 1} \CM_{1, 1}^{-1} \CM_{1, 2}}
\).
Blocking \( \CM \) such that 1 is the current column and 2 is
all the indices past 1, we observe the column of the Cholesky
factor is a conditional covariance \( \CM_{2, 1} \) divided by
\( \chol(\CM_{1, 1})^{\top} \), the square root of the variance
(the Cholesky factor of a scalar is its square root).
We conclude that the \( k \)-th column of the Cholesky factor equals \(
\vec{u} \) in \cref{eq:cond_select} since iteratively conditioning on the
columns \( i_1, i_2, \dotsc, i_{k - 1} \) is equivalent to conditioning
on \( \I \) by the quotient rule \cref{eq:quotient_rule}.

\subsection{Updating precision after insertion}
\label{app:prec_insert}

Given the precision matrix \( \CM_{1, 1}^{-1} \) we wish to
compute the inverse of the covariance matrix \( \CM_{1, 1} \)
with added rows and columns, i.e. to compute \( \CM^{-1} \) for
\(
  \CM =
    \begin{pmatrix}
      \CM_{1, 1} & \CM_{1, 2} \\
      \CM_{2, 1} & \CM_{2, 2}
    \end{pmatrix}
\).
Using the same block \( L D L^{\top} \) factorization as \cref{eq:chol} with
the Schur complement \textcolor{lightblue}{\( \CM_{2, 2} - \CM_{2, 1} \CM_{1,
1}^{-1} \CM_{1, 2} \)} denoted \textcolor{lightblue}{\( \CM_{2, 2 \mid 1} \)},
\begin{align}
  \begin{pmatrix}
    \CM_{1, 1} & \CM_{1, 2} \\
    \CM_{2, 1} & \CM_{2, 2}
  \end{pmatrix} &=
  \begin{pmatrix}
    \Id & 0 \\
    \textcolor{darkorange}{\CM_{2, 1} \CM_{1, 1}^{-1}} & \Id
  \end{pmatrix}
  \begin{pmatrix}
    \CM_{1, 1} & 0 \\
    0 & \textcolor{lightblue}{\CM_{2, 2 \mid 1}}
  \end{pmatrix}
  \begin{pmatrix}
    \Id & \textcolor{darkorange}{\CM_{1, 1}^{-1} \CM_{1, 2}} \\
    0 & \Id
  \end{pmatrix}.
  \shortintertext{Inverting both sides of the equation and multiplying,}
  \CM^{-1} &=
  \begin{pmatrix}
    \Id & -\textcolor{darkorange}{\CM_{1, 1}^{-1} \CM_{1, 2}} \\
    0 & \Id
  \end{pmatrix}
  \begin{pmatrix}
    \CM_{1, 1}^{-1} & 0 \\
    0 & \textcolor{lightblue}{\CM_{2, 2 \mid 1}^{-1}}
  \end{pmatrix}
  \begin{pmatrix}
    \Id & 0 \\
    -\textcolor{darkorange}{\CM_{2, 1} \CM_{1, 1}^{-1}} & \Id
  \end{pmatrix} \\
  &=
  \begin{pmatrix}
    \CM_{1, 1}^{-1} +
    \textcolor{darkorange}{
      \left (\CM_{1, 1}^{-1} \CM_{1, 2} \right )
    } \textcolor{lightblue}{\CM_{2, 2 \mid 1}^{-1}}
    \textcolor{darkorange}{
      \left (\CM_{2, 1} \CM_{1, 1}^{-1} \right )
    } &
    - \textcolor{darkorange}{
      \left (\CM_{1, 1}^{-1} \CM_{1, 2} \right )
    } \textcolor{lightblue}{\CM_{2, 2 \mid 1}^{-1}} \\
    - \textcolor{lightblue}{\CM_{2, 2 \mid 1}^{-1}} \textcolor{darkorange} {
      \left (\CM_{2, 1} \CM_{1, 1}^{-1} \right )
    } & \textcolor{lightblue}{\CM_{2, 2 \mid 1}^{-1}}
  \end{pmatrix}.
  \shortintertext{
    In the context of adding a new entry \( k \) to the
    matrix, \( \CM_{1, 1} = \CM_{\I, \I} \), \( \CM_{1, 2}
    = \CM_{\I, k} \), and \( \CM_{2, 2} = \CM_{k, k} \).
    Also note that \( \textcolor{lightblue}{\CM_{k, k \mid \I}^{-1}}
    \) is the precision of \( k \) conditional on the entries in \(
    \I \), which has already been computed in \cref{alg:select_prec}.
    If we let \( \vec{v} \defeq \textcolor{darkorange}{\CM_{\I,
    \I}^{-1} \CM_{\I, k}} \), then
  }
  &=
  \begin{pmatrix}
    \CM_{\I, \I}^{-1} + \CM_{k, k \mid \I}^{-1} \vec{v} \vec{v}^T &
    -\CM_{k, k \mid \I}^{-1} \vec{v} \\
    -\CM_{k, k \mid \I}^{-1} \vec{v}^{\top} & \CM_{k, k \mid \I}^{-1} \\
  \end{pmatrix}
\end{align}
which is precisely the update in line 13 of \cref{alg:select_prec}.
Note that the bulk of the update is a rank-one update to \( \CM_{1,
1}^{-1} \), which can be computed in \( \BigO(\card{\I}) = \BigO(s^2) \).

\subsection{Updating precision after conditioning}
\label{app:prec_cond}

Given the precision matrix of the prediction points conditional on the selected
entries, \( \CM_{\Pred, \Pred \mid \I}^{-1} \), we want to take into account
selecting an index \( k \), or to compute \( \CM_{\Pred, \Pred \mid \I, k}^{-1}
\), which is a rank-one update to the covariance matrix (but not necessarily
the precision matrix) from \cref{eq:cond_select}.
We can directly apply the Sherman–Morrison–Woodbury
formula which states that
\begin{align}
  \CM_{1, 1 \mid 2}^{-1} &= \CM_{1, 1}^{-1} +
    \left (\CM_{1, 1}^{-1} \CM_{1, 2} \right ) \CM_{2, 2 \mid 1}^{-1}
    \left (\CM_{2, 1} \CM_{1, 1}^{-1} \right ).
  \shortintertext{
    Expanding the Schur complement from the
    posterior covariance \cref{eq:cond_cov},
  }
  \left (
    \CM_{1, 1} - \CM_{1, 2} \CM_{2, 2}^{-1} \CM_{2, 1}
  \right )^{-1} &= \CM_{1, 1}^{-1} +
    \left (\CM_{1, 1}^{-1} \CM_{1, 2} \right ) \CM_{2, 2 \mid 1}^{-1}
    \left (\CM_{2, 1} \CM_{1, 1}^{-1} \right ).
  \shortintertext{
    For brevity of notation, define the vectors \( \vec{u}
    \defeq \CM_{1, 2} \) and \( \vec{v} \defeq \CM_{1,
    1}^{-1} \CM_{1, 2} = \CM_{1, 1}^{-1} \vec{u} \),
  }
  (\CM_{1, 1} - \CM_{2, 2}^{-1} \vec{u} \vec{u}^{\top})^{-1} &=
    \CM_{1, 1}^{-1} + \CM_{2, 2 \mid 1}^{-1} \vec{v} \vec{v}^{\top}.
  \shortintertext{
    So we see that a rank-one update to \( \CM_{1, 1} \) then
    inverting is a rank-one update to \( \CM_{1, 1}^{-1} \).
    In our context, \( \CM_{1, 1} = \CM_{\Pred, \Pred \mid \I}, \vec{u}
    = \CM_{\Pred, k \mid I}, \CM_{2, 2} = \CM_{k, k \mid \I} \) so \(
    \CM_{2, 2 \mid 1}^{-1} = \CM_{k, k \mid \Pred, I}^{-1} \) (from the
    quotient rule \cref{eq:quotient_rule}).
    By definition, \( \vec{v} = \CM_{\Pred, \Pred \mid \I}^{-1} \vec{u} \).
    Thus, the update in context is
  }
  \left ( \CM_{\Pred, \Pred \mid \I} -
    \frac{\CM_{\Pred, k \mid \I} \CM_{\Pred, k \mid \I}^{\top}}
         {\CM_{k, k \mid \I}}
  \right )^{-1} &=
    \CM_{1, 1}^{-1} +
    \CM_{k, k \mid \Pred, \I}^{-1} \vec{v} \vec{v}^{\top}
\end{align}
which is the update in line 18 of \cref{alg:select_mult_prec}.
Since the update is a rank-one update, it can be
computed in \( \BigO(\card{\Pred}^2) = \BigO(m^2) \).

\subsection{Updating log determinant after conditioning}
\label{app:logdet_downdate}

In the context of \cref{subsec:mult_select}, given the log determinant
of the covariance matrix of the prediction points conditional on the
selected entries, \( \logdet(\CM_{\Pred, \Pred \mid \I}) \), we wish to
compute the log determinant after we add an index \( k \) to \( \I \),
that is, to compute \( \logdet(\CM_{\Pred, \Pred \mid \I, k}) \).
\begin{proof}[Proof of Equation \cref{eq:greedy_mult}]
  From the posterior \cref{eq:cond_select}, selecting a new
  point is a rank-one downdate on the current covariance matrix,
  \begin{align}
    \nonumber
    \logdet(\CM_{\Pred, \Pred \mid \I, k})
    &= \logdet \left ( \CM_{\Pred, \Pred \mid \I} -
        \frac{\CM_{\Pred, k \mid \I} \CM_{\Pred, k \mid \I}^{\top}}
              {\CM_{k, k \mid \I}}
      \right ). \\
    \shortintertext{Applying the matrix determinant lemma,}
    &= \logdet(\CM_{\Pred, \Pred \mid \I}) +
      \log \left ( 1 -
        \frac{\CM_{\Pred, k \mid \I}^{\top} \CM_{\Pred, \Pred \mid \I}^{-1}
              \CM_{\Pred, k \mid \I}
              }{\CM_{k, k \mid \I}}
      \right ).
    \shortintertext{
      Focusing on the second term, we can turn
      the quadratic form into conditioning,
    }
    &= \logdet(\CM_{\Pred, \Pred \mid \I}) +
      \log \left (
        \frac{\CM_{k, k \mid \I} -
              \CM_{k, \Pred \mid \I} \CM_{\Pred, \Pred \mid \I}^{-1}
              \CM_{\Pred, k \mid \I}
              }{\CM_{k, k \mid \I}}
      \right ).
    \shortintertext{
      By the quotient rule \cref{eq:quotient_rule},
      we combine the conditioning,
    }
    &= \logdet(\CM_{\Pred, \Pred \mid \I}) +
      \log \left (
       \frac{\CM_{k, k \mid \I, \Pred}}{\CM_{k, k \mid \I}}
      \right ).
  \end{align}
\end{proof}


\section{Algorithms}

Reference Python and Cython code for all algorithms can be
found at \href{https://github.com/stephen-huan/conditional-knn}
{https://github.com/stephen-huan/conditional-knn}.

\subsection{Single-target selection}
\label{app:select}

To compute the objective \cref{eq:obj_gp} for each candidate index \( j \),
we start with the unconditional variance \( \CM_{j, j} \) and covariance \(
\CM_{\Pred, j} \), updating these quantities when an index \( k \) is selected.
If we can compute \( \vec{u} \) for \( k \), then we can update
\( j \)'s conditional variance by subtracting \( u_j^2 \) and
update its conditional covariance by subtracting \( u_j u_\Pred \).
We have two efficient strategies to compute \( \vec{u} \).

\paragraph{Precision approach}
(\cref{alg:select_prec}).

The precision of the selected entries \( \CM_{\I, \I}^{-1} \)
can be updated whenever a new index is added to \( \I \) in
time complexity \( \BigO(s^2) \) (see \cref{app:prec_insert}).
With \( \CM_{\I, \I}^{-1} \) in hand \( \vec{u} \) is computed
directly according to the posterior covariance \cref{eq:cond_cov}.
On each of the \( s \) rounds of selection it takes \( \BigO(s^2)
\) to update the precision and \( \BigO(Ns) \) to compute \(
\vec{u} \), for an overall time complexity of \( \BigO(N s^2) \).

\paragraph{Cholesky approach}
(\cref{alg:select_chol}).

The second strategy exploits the iterative conditioning in Cholesky
factorization: we maintain only the first \( \card{I} \) columns corresponding
to selected points from the Cholesky factor of the joint covariance matrix
between the training and target points; the \( k \)-th column of this factor
is precisely \( \vec{u} \) (see \cref{app:chol_stat} for details).
Adding a new column to the end of this factor can
be efficiently computed with left-looking in time complexity \( \BigO(N s) \)
(see \cref{alg:chol_update}), resulting in the same \( \BigO(N s^2) \) time
complexity as the explicit precision approach.

\subsection{Multiple-target selection}
\label{app:mult_select}

Context is provided in \cref{subsec:mult_select}.

\paragraph{Precision approach}
(\cref{alg:select_mult_prec}).

The precision approach initializes the precision \( \CM_{\Pred, \Pred}^{-1} \)
for \( m \) prediction points in time \( \BigO(m^3) \) and computes the initial
conditional variances \( \CM_{k, k \mid \Pred} \) for \( N \) candidates by
direct application of the posterior \cref{eq:cond_cov} in time \( \BigO(N m^2)
\).
For each of the \( s \) rounds of selecting candidates, it costs \(
\BigO(s^2) \) and \( \BigO(m^2) \) to update the precisions \( \CM_{\I,
\I}^{-1} \) and \( \CM_{\Pred, \Pred}^{-1} \) respectively, where the details
of efficiently updating \( \CM_{\Pred, \Pred}^{-1} \) after the rank-one
update are given in \cref{app:prec_cond}.
Given the precisions, \( \vec{u} \) \cref{eq:cond_select} and \( \vec{u}_\Pred
\defeq \frac{\CM_{:, k \mid \I, \Pred}}{\sqrt{\CM_{k, k \mid \I, \Pred}}} \)
are computed as usual according to \cref{eq:cond_cov} in time \( \BigO(N s) \)
and \( \BigO(N m) \).
Finally, for each candidate \( j \) the conditional variance \( \CM_{j, j \mid
\I} \) is updated by subtracting \( u_j^2 \), the conditional covariance \(
\CM_{\Pred, k \mid \I} \) is updated for each prediction point index \( c \)
by subtracting \( u_j u_c \), and the conditional variance \( \CM_{j, j \mid
I, \Pred} \) is updated by subtracting \( (u_\Pred)_j^2 \).
The total time complexity is \( \BigO(N s^2 + N m^2 + m^3) \).
The space complexity is \( \BigO(s^2 + m^2) \) to store both
precision matrices as well as \( \BigO(N m) \) memory to store
the conditional covariances of each candidate with each target.

\paragraph{Cholesky approach}
(\cref{alg:select_mult_chol}).

The approach storing two partial Cholesky factors is
considerably simpler than the explicit precision approach.
We first add each prediction point to one Cholesky factor by
left-looking (\cref{alg:chol_update}) exactly as in the single-target
algorithm, for a time complexity of \( \BigO((N + m) m^2) \).
Whenever a candidate is selected, both Cholesky factors are
updated in time \( \BigO((N + m)(m + s)) \) dominated by
updating the Cholesky factor conditioned on prediction points.
The desired covariances \( \vec{u} \) and \( \vec{u}_\Pred \) are directly
read off the columns of the Cholesky factors and both conditional variances
\( \CM_{j, j \mid I} \) and \( \CM_{j, j \mid I, \Pred} \) are computed like
in the precision approach.
For \( s \) points selected, the total time
complexity is \( \BigO(N s^2 + N m^2 + m^3) \).
The space complexity is \( \BigO(N s + N m + m^2) \) to store two
\( (N + m) \times (m + s) \) sized partial Cholesky factors for
the joint covariance matrix of training and prediction points.


\subsection{Partial selection}
\label{subsec:partial_select}

We now compute the log determinant of the target covariance
matrix if a selected index \( k \) only conditions a subset
of the target points, skipping the first \( p \) targets.
Following the suggestive form of the target covariance matrix
\cref{eq:chol_partial}, we maintain a Cholesky factor \( L \)
taking from the factor of the joint covariance matrix \( \CM \)
only the \( \card{I} + m \) columns corresponding to selected
points and target points, sorted w.r.t.\ \( \succ \).
After inserting the index \( k \) into its proper
column in \( L \), the updated \( L \) becomes
\begin{align}
  \label{eq:chol_partial_update}
  L &\gets
  \begin{pmatrix}
    L_{:, :p} & \vec{u} & {L'}_{:, p + 1:}
  \end{pmatrix} &
    \vec{u} &= \frac{\CM_{:, k \mid :p}}{\sqrt{\CM_{k, k \mid :p}}}
\end{align}
where \( L' \) is the Cholesky factor of the covariance matrix
conditional on \( k \) and the updated form of \( L \) after insertion
is derived from the statistical perspective in \cref{eq:chol}.

To compute this insertion, we immediately see that the
first \( p \) columns \( L_{:, :p} \) are unchanged.
We compute the newly inserted column \( \vec{u} \) of the form
\cref{eq:cond_select} as usual with left-looking (see \cref{alg:chol_update}).
Finally, we can efficiently compute the remaining columns \( {L'}_{:, p +
1:} \) from the original factor \( L_{:, p + 1:} \) by observing that the
updated factor \( {L'}_{:, p + 1:} \) is identical to the Cholesky factor
of \( \CM \) after a rank-one downdate by \( \vec{u} \).
Specifically we adapt Lemma 1 of \cite{krause2015more} to make
no assumption on the row ordering of \( L \) in order to
implement the downdate in-place (see \cref{alg:chol_insert}).
For a Cholesky factor of size \( R \times C \), the time
complexity of the downdate is \( \BigO(R C) \) compared
to \( \BigO(R C^2) \) if recomputed from scratch.

For \( N \) candidates, \( m \) targets, and a budget of \( s \) points to
select, the Cholesky factor \( L \) has size \( (N + m) \times (s + m) \).
When adding a new index \( k \) that ignores the first \( p \) targets,
the first \( p \) columns of \( L \) are correspondingly ignored.
We compute the new column \( \vec{u} \) with
left-looking in time \( \BigO((N + m) p) \).
Finally, we update the columns past \( p \) with
downdating in time \( \BigO((N + m) (s + m - p)) \) for
a total time complexity of \( \BigO(s(N + m)(s + m)) \).

With the factor \( L \) in hand, we discuss
actually computing the log determinant objective
\cref{eq:partial_kl} after adding a candidate index \( j \).
We directly compute each individual target point's variance,
conditional on \( j \) as well as previously selected points, and
simply add these variances together to get the overall objective.

\newpage

We proceed by induction on the columns of \( L \).
Starting at the first column, we know the candidate's unconditional variance.
Assuming we know the candidate's variance conditional on all points prior,
we proceed to the next column by reading the requisite
statistical quantities from \( L \) \cref{eq:chol_partial_update}
and conditioning \cref{eq:quotient_rule}.
For the \( i \)-th column,
\begin{align}
  \label{eq:partial_diag}
  \CM_{i, i \mid :i - 1} &= L_{i, i}^2 \\
  \CM_{j, i \mid :i - 1} &= L_{j, i} \cdot L_{i, i} \\
  \label{eq:obj_partial}
  \CM_{i, i \mid :i - 1, j} &= \CM_{i, i \mid :i - 1} -
    \CM_{j, i \mid :i - 1}^2/\CM_{j, j \mid :i - 1} \\
  \label{eq:partial_induct}
  \CM_{j, j \mid :i - 1, i} &= \CM_{j, j \mid :i - 1} -
    \CM_{j, i \mid :i - 1}^2/\CM_{i, i \mid :i - 1} = \CM_{j, j \mid :i}
\end{align}
where \cref{eq:partial_induct} satisfies the conditions of the inductive
hypothesis for the next column and \cref{eq:obj_partial} is the
desired conditional variance when the \( i \)-th point is a target.

For each of the \( N \) candidates, it requires \( \BigO(1) \) work per column
for \( s + m \) columns. Over \( s \) selections, the total time complexity is
\( \BigO(s N (s + m)) \) which is dominated by the time to update the Cholesky
factor, meaning the partial selection algorithm matches the asymptotic time
complexity of the multiple-target algorithm.
However, the downdate of the Cholesky factor is implemented with
repeated BLAS level-one \texttt{daxpy} operations while the bulk of
left-looking takes place in a BLAS level-two \texttt{dgemv} operation.
Higher-level operations often have better constant-factor
performance for the same asymptotic time complexity.
In addition, the objective for the partial selection algorithm is
more complicated to compute than its multiple-target counterpart.
Pseudocode for the algorithm is provided in
\cref{alg:select_partial} to \cref{alg:chol_insert}.

\begin{figure}[th!]
  \centering
  \begin{minipage}[t]{0.49\textwidth}
    \begin{algorithm}[H]
      \caption{Point selection \\ by explicit precision}
      \label{alg:select_prec}
      \input{figures/algorithms/select_prec.tex}
    \end{algorithm}
  \end{minipage}
  \hfill
  \begin{minipage}[t]{0.49\textwidth}
    \begin{algorithm}[H]
      \caption{Point selection \\ by Cholesky factorization}
      \label{alg:select_chol}
      \input{figures/algorithms/select_chol.tex}
    \end{algorithm}
  \end{minipage}
  \caption{Algorithms for single-target selection.}
  \label{fig:alg_select}
\end{figure}

\begin{algorithm}[H]
  \caption{Directed sparse Gaussian process regression by selection}
  \label{alg:infer_select}
  \begin{algorithmic}[1]
  \REQUIRE \(
    X_\Train, \vec{y}_\Train,
    X_\Pred, \K(\cdot, \cdot), s
  \)
  \ENSURE \(
      \E[\vec{y}_\Pred \mid \vec{y}_\Train],
    \Cov[\vec{y}_\Pred \mid \vec{y}_\Train]
  \)

  \STATE Compute \( \I \) using
    \cref{alg:select_prec} or \cref{alg:select_chol}.
  \STATE \(
    \CM_{\Train, \Train} \gets
    \K(X_\Train[I], X_\Train[\I])
  \)
  \STATE \(
    \CM_{\Pred, \Pred} \gets
    K(X_\Pred, X_\Pred)
  \)
  \STATE \(
    \CM_{\Train, \Pred} \gets
    K(X_\Train[\I], X_\Pred)
  \)
  \STATE \(
    \E[\vec{y}_\Pred \mid \vec{y}_\Train] \gets
    \CM_{\Pred, \Train} \Theta_{\Train, \Train}^{-1}
    \vec{y}_\Train[\I]
  \)
  \STATE \(
    \Cov[\vec{y}_\Pred \mid \vec{y}_\Train] \gets
    \CM_{\Pred, \Pred} -
    \CM_{\Train, \Pred}^{\top} \Theta_{\Train, \Train}^{-1}
    \CM_{\Train, \Pred}
  \)
  \RETURN \(
      \E[\vec{y}_\Pred \mid \vec{y}_\Train],
    \Cov[\vec{y}_\Pred \mid \vec{y}_\Train]
  \)
\end{algorithmic}

\end{algorithm}

\clearpage

\begin{figure}[H]
  \centering
  \begin{minipage}[t]{0.49\textwidth}
    \begin{algorithm}[H]
      \caption{Multiple-target selection \\ by explicit precision}
      \label{alg:select_mult_prec}
      \input{figures/algorithms/select_mult_prec.tex}
    \end{algorithm}
  \end{minipage}
  \hfill
  \begin{minipage}[t]{0.49\textwidth}
    \begin{algorithm}[H]
      \caption{Multiple-target selection \\ by Cholesky factorization}
      \label{alg:select_mult_chol}
      \input{figures/algorithms/select_mult_chol.tex}
    \end{algorithm}

    \begin{algorithm}[H]
      \caption{Update Cholesky factor}
      \label{alg:chol_update}
      \input{figures/algorithms/chol_update.tex}
    \end{algorithm}
  \end{minipage}
  \caption{Algorithms for multiple-target selection.}
  \label{fig:alg_mult_select}
\end{figure}

\begin{figure}[H]
  \centering
  \begin{minipage}[t]{0.49\textwidth}
    \begin{algorithm}[H]
      \caption{Partial point selection}
      \label{alg:select_partial}
      \input{figures/algorithms/select_partial.tex}
    \end{algorithm}

    \begin{algorithm}[H]
      \caption{Find \( j \)'s index in \( \Order \)}
      \label{alg:insert_index}
      \input{figures/algorithms/insert_index.tex}
    \end{algorithm}
  \end{minipage}
  \hfill
  \begin{minipage}[t]{0.49\textwidth}
    \begin{algorithm}[H]
      \caption{Compute \( j \)'s objective}
      \label{alg:partial_score}
      \input{figures/algorithms/partial_score.tex}
    \end{algorithm}
    \vspace{-7pt}
    \begin{algorithm}[H]
      \caption{Insert \( k \) into \( L \)}
      \label{alg:chol_insert}
      \input{figures/algorithms/chol_insert.tex}
    \end{algorithm}
  \end{minipage}
  \caption{Algorithms for partial selection.}
  \label{fig:alg_partial_select}
\end{figure}

\newpage

\section{Additional figures}

\subsection{Additional numerical results}

The following plots \cref{fig:chol_n_norm}, \cref{fig:chol_rho_norm},
and \cref{fig:chol_time_norm} are variants of \cref{fig:chol_n}
and \cref{fig:chol_rho} for the Frobenius norm \( \norm{\CM -
(L L^\top)^{-1}}_{\FRO} \) and operator norm \( \norm{\CM - (L
L^\top)^{-1}}_2 \) as the error metric instead of the KL divergence.
The plot \cref{fig:cg_norm} is a variant of \cref{fig:cg_rho}
showing \( \norm{\Id - \CM L L^\top}_2 \) for each preconditioner.

\pgfplotsset{
  cycle list={
    {very thick, lightblue, style=dashed,         mark=*},
    {very thick, seagreen,  style=dashdotted,     mark=o},
    {very thick, silver,    style=densely dotted, mark=triangle*},
    {very thick, orange,    style=solid,          mark=square*},
    {very thick, rust,      style=dotted,         mark=square},
    {very thick, joshua,    style=loosely dashed, mark=diamond*},
  }
}

\begin{figure}[h]
  \centering
  \begin{tikzpicture}[baseline, scale=0.60]
  \begin{axis}[
    grid={major},
    xlabel={\( N \)},
    ylabel={Frobenius norm \( \norm{\CM - (L L^\top)^{-1}}_{\FRO} \)},
    legend entries={%
      \( \rho \)-ball, \( \rho \)-ball (agg.),
      \( k \)-NN, select, select (agg.)
    },
    legend pos={north west},
  ]
  \addplot table {figures/cholesky_norm/data/n_fro_norm_KL.csv};
  \addplot table {figures/cholesky_norm/data/n_fro_norm_KL_agg.csv};
  \addplot table {figures/cholesky_norm/data/n_fro_norm_select-KNN.csv};
  \addplot table {figures/cholesky_norm/data/n_fro_norm_select.csv};
  \addplot table {figures/cholesky_norm/data/n_fro_norm_select_agg.csv};
  \end{axis}
\end{tikzpicture}%
  \begin{tikzpicture}[baseline, scale=0.60]
  \begin{axis}[
    grid={major},
    xlabel={\( N \)},
    ylabel={Operator norm \( \norm{\CM - (L L^\top)^{-1}}_2 \)},
    legend pos={north west},
  ]
  \addplot table {figures/cholesky_norm/data/n_op_norm_KL.csv};
  \addplot table {figures/cholesky_norm/data/n_op_norm_KL_agg.csv};
  \addplot table {figures/cholesky_norm/data/n_op_norm_select-KNN.csv};
  \addplot table {figures/cholesky_norm/data/n_op_norm_select.csv};
  \addplot table {figures/cholesky_norm/data/n_op_norm_select_agg.csv};
  \end{axis}
\end{tikzpicture}
  \caption{%
    Frobenius norm (left) and operator norm (right) of
    Cholesky factorization methods with varying number
    of points \( N \) and fixed density \( \rho = 2 \).
  }
  \label{fig:chol_n_norm}
\end{figure}

\begin{figure}[h]
  \centering
  \begin{tikzpicture}[baseline, scale=0.60]
  \begin{semilogyaxis}[
    grid={major},
    xlabel={Density (\( \mathsf{nnz} / N^2 \))},
    ylabel={Frobenius norm \( \norm{\CM - (L L^\top)^{-1}}_{\FRO} \)},
    legend entries={%
      \( \rho \)-ball, \( \rho \)-ball (agg.),
      \( k \)-NN, select, select (agg.)
    },
    legend pos={north east},
  ]
  \addplot table {figures/cholesky_norm/data/rho_fro_norm_KL.csv};
  \addplot table {figures/cholesky_norm/data/rho_fro_norm_KL_agg.csv};
  \addplot table {figures/cholesky_norm/data/rho_fro_norm_select-KNN.csv};
  \addplot table {figures/cholesky_norm/data/rho_fro_norm_select.csv};
  \addplot table {figures/cholesky_norm/data/rho_fro_norm_select_agg.csv};
  \end{semilogyaxis}
\end{tikzpicture}%
  \begin{tikzpicture}[baseline, scale=0.60]
  \begin{semilogyaxis}[
    grid={major},
    xlabel={Density (\( \mathsf{nnz} / N^2 \))},
    ylabel={Operator norm \( \norm{\CM - (L L^\top)^{-1}}_2 \)},
    legend pos={north east},
  ]
  \addplot table {figures/cholesky_norm/data/rho_op_norm_KL.csv};
  \addplot table {figures/cholesky_norm/data/rho_op_norm_KL_agg.csv};
  \addplot table {figures/cholesky_norm/data/rho_op_norm_select-KNN.csv};
  \addplot table {figures/cholesky_norm/data/rho_op_norm_select.csv};
  \addplot table {figures/cholesky_norm/data/rho_op_norm_select_agg.csv};
  \end{semilogyaxis}
\end{tikzpicture}
  \caption{%
    Frobenius norm (left) and operator norm (right) with varying
    density \( \rho \). The number of points is \( N = 2^{16} \).
  }
  \label{fig:chol_rho_norm}
\end{figure}

\begin{figure}[h]
  \centering
  \begin{tikzpicture}[baseline, scale=0.60]
  \begin{loglogaxis}[
    grid={major},
    xlabel={Time (seconds)},
    ylabel={Frobenius norm \( \norm{\CM - (L L^\top)^{-1}}_{\FRO} \)},
    legend entries={%
      \( \rho \)-ball, \( \rho \)-ball (agg.),
      \( k \)-NN, select, select (agg.)
    },
    legend pos={north east},
  ]
  \addplot table {figures/cholesky_norm/data/rho_time_KL_fro_norm.csv};
  \addplot table {figures/cholesky_norm/data/rho_time_KL_agg_fro_norm.csv};
  \addplot table {figures/cholesky_norm/data/rho_time_select-KNN_fro_norm.csv};
  \addplot table {figures/cholesky_norm/data/rho_time_select_fro_norm.csv};
  \addplot table {figures/cholesky_norm/data/rho_time_select_agg_fro_norm.csv};
  \end{loglogaxis}
\end{tikzpicture}%
  \begin{tikzpicture}[baseline, scale=0.60]
  \begin{loglogaxis}[
    grid={major},
    xlabel={Time (seconds)},
    ylabel={Operator norm \( \norm{\CM - (L L^\top)^{-1}}_2 \)},
  ]
  \addplot table {figures/cholesky_norm/data/rho_time_KL_op_norm.csv};
  \addplot table {figures/cholesky_norm/data/rho_time_KL_agg_op_norm.csv};
  \addplot table {figures/cholesky_norm/data/rho_time_select-KNN_op_norm.csv};
  \addplot table {figures/cholesky_norm/data/rho_time_select_op_norm.csv};
  \addplot table {figures/cholesky_norm/data/rho_time_select_agg_op_norm.csv};
  \end{loglogaxis}
\end{tikzpicture}
  \caption{%
    Accuracy in Frobenius norm (left) and operator norm
    (right) to computational time trade-off over varying
    \( \rho \). The number of points is \( N = 2^{16} \).
  }
  \label{fig:chol_time_norm}
\end{figure}

\clearpage

\begin{figure}[t]
  \centering
  \begin{tikzpicture}[baseline, scale=0.75]
  \begin{loglogaxis}[
    grid={major},
    xlabel={Density (\( \mathsf{nnz} / N^2 \))},
    ylabel={Operator norm \( \norm{\Id - \CM L L^\top} \)},
    legend entries={%
      \( \rho \)-ball, \( \rho \)-ball (agg.),
      \( k \)-NN, select, select (agg.), hlib
    },
    legend pos={south west},
  ]
  \addplot table {figures/cg_norm/data/rho_op_norm_KL.csv};
  \addplot table {figures/cg_norm/data/rho_op_norm_KL_agg.csv};
  \addplot table {figures/cg_norm/data/rho_op_norm_select-KNN.csv};
  \addplot table {figures/cg_norm/data/rho_op_norm_select.csv};
  \addplot table {figures/cg_norm/data/rho_op_norm_select_agg.csv};
  \addplot table {figures/cg_norm/data/rho_op_norm_hlib.csv};
  \end{loglogaxis}
\end{tikzpicture}%
  \caption{%
    Operator norm of various preconditioners for \( N = 2^{17} \) points.
  }
  \label{fig:cg_norm}
\end{figure}

\subsection{Conditional point selection}

The following figures are variations of \cref{fig:cond_select} for
different point geometries. Like \cref{fig:cond_select}, the first
panel shows selecting the \( k \)-\textcolor{seagreen}{nearest
neighbors} to the \textcolor{orange}{center point} out of many possible
\textcolor{silver}{candidates}. The next panels show selection by greedily
maximizing conditional mutual information with the center point for a
Mat{\'e}rn kernel with length scale \( \ell = 1 \) and increasing smoothness
\( \nu \), from left to right: \( \nu = 1/2, 3/2, 5/2 \).

\begin{figure}[H]
  \centering
  \includegraphics{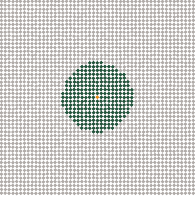}%
  \quad
  \includegraphics{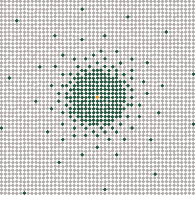}%
  \quad
  \includegraphics{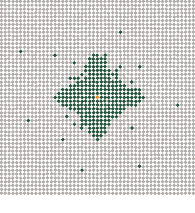}%
  \quad
  \includegraphics{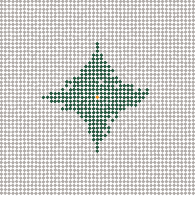}%
  \caption{%
    Dense grid with a random perturbation
    uniformly sampled from \( \pm 10^{-3} \).
  }
\end{figure}

\begin{figure}[H]
  \centering
  \includegraphics{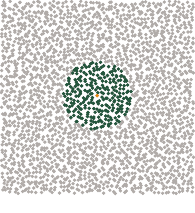}%
  \quad
  \includegraphics{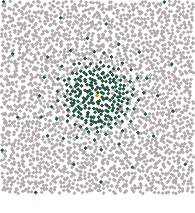}%
  \quad
  \includegraphics{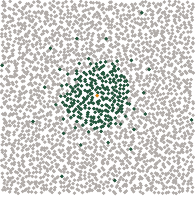}%
  \quad
  \includegraphics{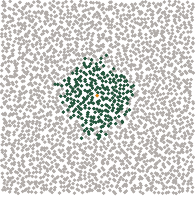}%
  \caption{%
    Dense grid with a random perturbation
    uniformly sampled from \( \pm 10^{-2} \).
  }
\end{figure}

\begin{figure}[H]
  \centering
  \includegraphics{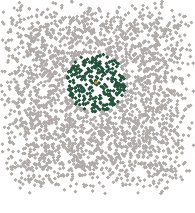}%
  \quad
  \includegraphics{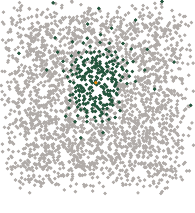}%
  \quad
  \includegraphics{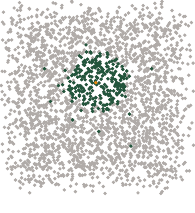}%
  \quad
  \includegraphics{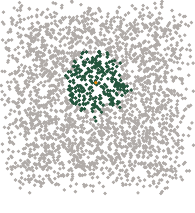}%
  \caption{%
    Dense grid with a random perturbation
    uniformly sampled from \( \pm 10^{-1} \).
  }
\end{figure}

\begin{figure}[H]
  \centering
  \includegraphics{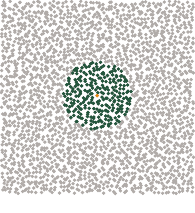}%
  \quad
  \includegraphics{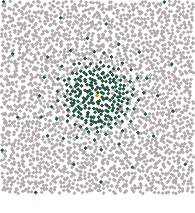}%
  \quad
  \includegraphics{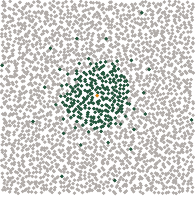}%
  \quad
  \includegraphics{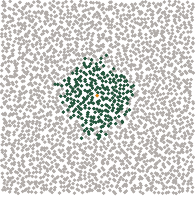}%
  \caption{%
    First points taken from a maximin ordering
    on a perturbed grid at \( \pm 10^{-5} \).
  }
\end{figure}

\begin{figure}[H]
  \centering
  \includegraphics{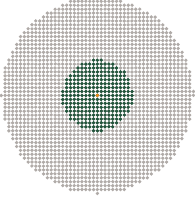}%
  \quad
  \includegraphics{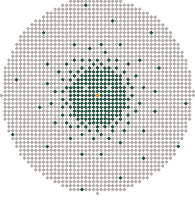}%
  \quad
  \includegraphics{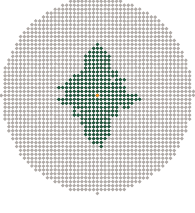}%
  \quad
  \includegraphics{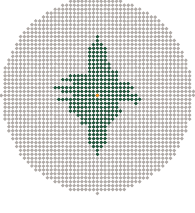}%
  \caption{%
    Points taken from a regular grid intersected
    with a circular domain and noise \( 10^{-3} \).
  }
\end{figure}

\begin{figure}[H]
  \centering
  \includegraphics{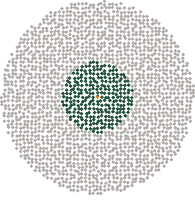}%
  \quad
  \includegraphics{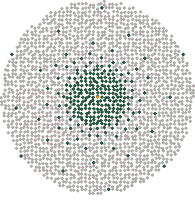}%
  \quad
  \includegraphics{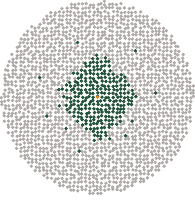}%
  \quad
  \includegraphics{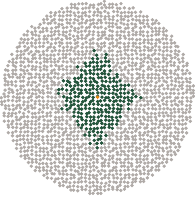}%
  \caption{%
    Points taken from a regular grid intersected
    with a circular domain and noise \( 10^{-2} \).
  }
\end{figure}

\begin{figure}[H]
  \centering
  \includegraphics{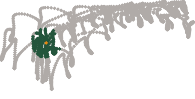}%
  \quad
  \includegraphics{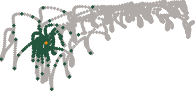}%
  \quad
  \includegraphics{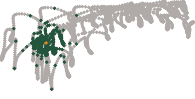}%
  \quad
  \includegraphics{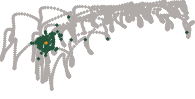}%
  \caption{%
    Points taken from the SARCOS dataset used in \cref{subsec:gp_exp}.
    Note that conditional selection aborts early if the decrease in
    variance is sufficiently small, so there may not be the same number
    of points selected between the panels. In particular, the highest
    smoothness (rightmost) panel selects fewer points.
  }
\end{figure}

\begin{figure}[H]
  \centering
  \includegraphics{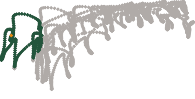}%
  \quad
  \includegraphics{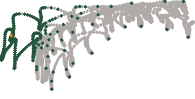}%
  \quad
  \includegraphics{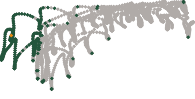}%
  \quad
  \includegraphics{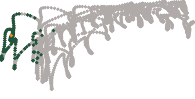}%
  \caption{%
    Points taken from the SARCOS dataset.
  }
\end{figure}

\begin{figure}[H]
  \centering
  \includegraphics{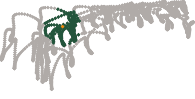}%
  \quad
  \includegraphics{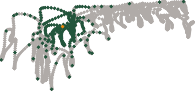}%
  \quad
  \includegraphics{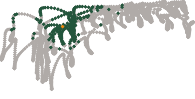}%
  \quad
  \includegraphics{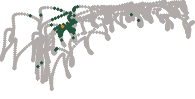}%
  \caption{%
    Points taken from the SARCOS dataset.
  }
\end{figure}

\end{document}